\documentclass[sigconf,authorversion,10pt,nonacm,natbib=false]{acmart}
\AtBeginDocument{%
  }

\setcopyright{acmlicensed}
\copyrightyear{2018}
\acmYear{2018}
\acmDOI{XXXXXXX.XXXXXXX}
\acmConference[Conference acronym 'XX]{Make sure to enter the correct
  conference title from your rights confirmation email}{June 03--05,
  2018}{Woodstock, NY}
\acmISBN{978-1-4503-XXXX-X/2018/06}

\RequirePackage[
  datamodel=acmdatamodel,
  style=acmnumeric,
  ]{biblatex}

\usepackage{amsmath}
\usepackage{multirow}
\usepackage[most]{tcolorbox}
\usepackage{dsfont}

\usepackage{subcaption}
\usepackage{graphicx}
\usepackage{algorithm}
\usepackage{algorithmic}
\usepackage{makecell}

\usepackage{xspace}
\newcommand{\sys}{\textsc{Cleave}\xspace}

\newcommand{\tinyskip}{\vspace{3pt}}
\newcommand{\mypar}[1]{\tinyskip\noindent\textbf{#1.}\xspace}

\newcommand*\myc[1]{%
\scalebox{0.78}{\begin{tikzpicture}[baseline=-3pt]
  \node[draw,circle,inner sep=0.5pt, fill=black] {\textcolor{white}{\textsf{\textbf{#1}}}};
\end{tikzpicture}}}

\usepackage{comment}

\usepackage{mathtools}
\newcommand{\eg}{\text{e.g.,}\ }
\newcommand{\st}{\text{s.t.}\ }
\newcommand{\ie}{\text{i.e.,}\ }

\newcommand{\R}{\mathbb{R}}

\newenvironment{tightlist}{
\begin{list}{$\bullet$}{
    \setlength{\topsep}{.1em}
    \setlength{\partopsep}{0in}
    \setlength{\parskip}{0in}
    \setlength{\itemsep}{0in}
    \setlength{\parsep}{0in}
    \setlength{\leftmargin}{1em}
    \setlength{\rightmargin}{0in}
    \setlength{\itemindent}{0in}
}}
{\end{list}}

\newcommand*{\RELEASE}{}
\ifdefined\RELEASE
  \newcommand{\review}[1]{\textcolor{black}{#1}}
  \newcommand{\ignore}[1]{}
  \newcommand{\fixme}[1]{}
  \newcommand{\lee}[1]{}
  \newcommand{\leyang}[1]{}
  \newcommand{\TODO}[1]{}
  
\else
  \newcommand{\ignore}[1]{}
  \newcommand{\fixme}[1]{{\textcolor{red}{[~FIXME:~#1~]}}}
  \newcommand{\lee}[1]{{\textcolor{orange}{[~LEE:~#1~]}}}
  \newcommand{\leyang}[1]{{\noindent\textcolor{blue}{[LX:~#1]}}}
  \newcommand{\TODO}[1]{{\textcolor{red}{TODO:~#1}}}
  
  \newcommand{\review}[1]{\textcolor{blue}{#1}}
\fi

\newcommand{\algorithmstyle}[1]{\renewcommand{\algocf@style}{#1}}

\newlength\myindent
\newcommand{\xmark}{%
\tikz[scale=0.23] {
    \draw[line width=0.7,line cap=round] (0,0) to [bend left=6] (1,1);
    \draw[line width=0.7,line cap=round] (0.2,0.95) to [bend right=3] (0.8,0.05);
}}
\newcommand{\cmark}{%
\tikz[scale=0.23] {
    \draw[line width=0.7,line cap=round] (0.25,0) to [bend left=10] (1,1);
    \draw[line width=0.8,line cap=round] (0,0.35) to [bend right=1] (0.23,0);
}}

\begin{document}

\title{On Harnessing Idle Compute at the Edge for Foundation Model Training}

\author{Leyang Xue$^\dag$, Meghana Madhyastha$^\ddag$, Myungjin Lee$^\diamond$, Amos Storkey$^\dag$, Randal Burns$^\ddag$ and Mahesh K. Marina$^\dag$}

\affiliation{%
	\institution{The University of Edinburgh$^\dag$ Johns Hopkins University$^\ddag$ Cisco Research$^\diamond$}
	\country{}
}

\begin{abstract}
	The foundation-model ecosystem remains highly centralized because training requires immense compute resources and is therefore largely limited to large cloud operators.
	Edge-assisted foundation model training that harnesses spare compute on edge devices offers a more democratized alternative.
	However, existing edge-training approaches fall short: they struggle to match cloud-training performance, scale to larger models, fit within device memory limits, or keep communication overhead manageable.
	They also do not handle device heterogeneity and churn satisfactorily.

	\review{We introduce \sys, built on a structural insight: each GEMM has an asymmetric I/O pattern---its input matrices, sent over downlink, are much larger than the partial output blocks returned over uplink---matching edge networks where downlink bandwidth exceeds uplink by 2--10$\times$. Exploiting this alignment with a parameter-server-centric architecture, \sys makes per-device communication \emph{decrease} as more devices join, rather than stay constant as in conventional TP. Decomposing training into independent sub-GEMM tasks yields one scheduling abstraction that unifies memory constraints, communication overhead, and fault tolerance under device churn.}

	Our evaluation shows that \sys achieves cloud-comparable GPU training performance and outperforms state-of-the-art edge-training methods by 4--10$\times$ in per-batch runtime at the same device counts. Beyond this shared operating range, \sys scales to thousands of heterogeneous devices---a regime where prior edge-training systems cannot operate---and achieves at least $100\times$ faster recovery from device failures.
\end{abstract}

\maketitle

\begin{tcolorbox}[colback=yellow!10, colframe=black, boxrule=0.5pt,
		arc=2pt, left=4pt, right=4pt, top=4pt, bottom=4pt, enhanced,
		sharp corners, width=\columnwidth, boxsep=2pt]
	\small
	\textbf{Note:} An extended abstract of this paper appeared in ACM MobiCom 2025. A workshop version appeared in EuroMLSys 2026 (co-located with EuroSys).
\end{tcolorbox}


\section{Introduction}

``Foundation models'' (FMs)~\cite{DBLP:journals/corr/abs-2108-07258} are driving the AI revolution~\cite{microsoft-ai-market}, enabling powerful generative systems across language~\cite{gpt4}, vision~\cite{stable-diffussion}, networks~\cite{netllm}, and software~\cite{copilot}.
Training these models on large-scale data for diverse tasks has intensified concerns about the centralization of the current ecosystem~\cite{Bartoldson24}.
This centralization stems from the immense compute resources and economic cost required for training~\cite{large-model-cost}, which only a handful of global entities---mostly cloud data-center operators---can afford.
As a response to such centralization, edge-assisted foundation model training has emerged~\cite{learning@home,dedloc}, leveraging volunteered edge devices to pool compute resources. This paradigm taps into the scale and underutilization of modern edge hardware~\cite{asteroid}, offering a more inclusive alternative to centralized training.

Edge-assisted foundation model training with edge devices would be appealing only if it satisfies \textbf{three key requirements}:
\textbf{First}, it should offer per-batch training time comparable to current cloud-based training systems.
\textbf{Second}, it should match the accuracy of training in the cloud by allowing the use of the same model architecture, optimizers and training hyperparameters (\eg batch size, sequence length).
This rules out using typical edge-device-oriented optimizations such as model or gradient compression, since those risk accuracy loss.
\textbf{Third}, it must support training large models on foundation-scale datasets, in line with the scaling laws that underpin the performance of modern neural architectures~\cite{scalelaw}.

However, current edge-training methods do not satisfy these requirements, which limits the practical appeal of this paradigm.
The problem is not a single missing optimization, but a mismatch between existing parallelization strategies and the constraints of edge environments.
In practice, this mismatch appears along \textbf{three coupled challenge axes}:

\mypar{1. Excessive per-device memory consumption} Existing edge-training approaches (\eg DTFM~\cite{dtfm}, EDDL~\cite{eddl}) employ data parallelism (DP), pipeline parallelism (PP), or a combination of both.
While the typical usable memory on phones is around 512MB~\cite{flexnn}, the memory demand of training a model can be hundreds of GB.
Even with the forms of model parallelism commonly adopted in edge training, the per-device memory demand remains too high across model sizes (detailed in \S\ref{sec:memory-demand}).

\mypar{2. High communication volume among devices}
Although tensor parallelism (TP) used in cloud training approaches (\eg Alpa~\cite{alpa}) is a key enabler for reducing per-device memory consumption, it shifts the problem to communication.
In addition to gradient AllReduce in DP, TP introduces additional AllReduce and AlltoAll at each layer in both backward and forward propagation (detailed in \S\ref{sec:memory-demand}).
Naively using TP to fit training within device capabilities therefore creates high communication overhead, making it difficult to use in edge settings with constrained network bandwidth.


\mypar{3. Handling device heterogeneity and churn}
Current approaches (\eg DTFM, Alpa) involve \emph{all} devices in training without fully considering their heterogeneity in compute and communication characteristics.
This ends up including stragglers in the communication-intensive DP and TP, slowing down the training process.
Moreover, edge environments are inherently dynamic---devices can disconnect, fail, or join at any point during training.
Current approaches either assume a static device set (\eg Alpa, DTFM) or lack fine-grained fault recovery and seamless integration of newly available devices (\eg SWARM~\cite{swarm}, Asteroid~\cite{asteroid}), making them inefficient in real-world edge deployments.

We present \sys, a parameter--server-centric framework that makes tensor-parallel FM training practical on heterogeneous edge devices.
\review{Building on the observation established in \S\ref{sec:background} that FM training is dominated by GEMM operations, our key insight is that each GEMM exhibits an asymmetric I/O pattern: the input matrices (distributed to devices over downlink) are substantially larger than the per-device partial output (returned over uplink). This structure \emph{aligns} with edge network links where downlink bandwidth exceeds uplink by 2--10$\times$. A parameter-server-centric architecture directly exploits this alignment: the PS dispatches row/column shards over downlink and collects small partial outputs over uplink, making per-device communication \emph{decrease} as the device count grows---the ideal scaling behavior in Figure~\ref{fig:comm-volume-problem}. Moreover, decomposing training into independent sub-GEMM tasks yields a unified scheduling abstraction that simultaneously addresses device memory limits (each device holds only its assigned shards), communication overhead (total GEMM volume is bounded, so per-device share shrinks with scale), and fault tolerance (a failed device loses only its shards, redistributed via the same cost model).}

Specifically, we make four contributions:

\begin{tightlist}
	\item~\textbf{\emph{Contribution \#1:}} \review{We identify and exploit the structural alignment between GEMM I/O asymmetry and edge network link asymmetry. By sharding each GEMM into row/column sub-tasks dispatched over downlink and collected over uplink, \sys achieves per-device communication that \emph{decreases} with scale---the ideal behavior that naive tensor parallelism cannot reach (\S\ref{sec:obervations}).}
	\item~\textbf{\emph{Contribution \#2:}} \review{We design a parameter-server-centric training framework (\S\ref{sec:overview}) that structurally exploits GEMM I/O asymmetry: the PS serves as the aggregator for all intermediate results, replacing peer-to-peer AllReduce and AlltoAll with a single downlink dispatch and uplink collection. This removes peer-to-peer collectives and directly leverages the higher downlink capacity typical of edge networks.}
	\item~\textbf{\emph{Contribution \#3:}} We design a cost model that selects devices for training and distributes workload among them (\S\ref{sec:cost-model}). The model accounts for compute and communication heterogeneity, as well as device churn. It enables stragglers to be used sparingly or excluded entirely by redistributing their workload to other devices, thereby avoiding stalls in the training process. It also leverages fine-grained sharding with TP and DP to ensure fast recovery under churn.
	\item~\textbf{\emph{Contribution \#4:}}
	We evaluate \sys against state-of-the-art (SOTA) edge (DTFM~\cite{dtfm}) and cloud (Alpa~\cite{alpa}) methods by training two foundation language-model families of various sizes---OPT~\cite{opt} and Llama2~\cite{llama2} (\S\ref{sec:evaluation}).
	In our evaluated setting, \sys achieves cloud-comparable per-batch runtime while supporting larger models than edge baselines, reduces per-batch runtime by $4$--$10\times$, and scales to thousands of devices ($2$--$8\times$ more than baselines).
	Within the device-count range where both \sys and the baselines operate (32--512 devices), \sys achieves 4--10$\times$ lower per-batch runtime (Figure~\ref{fig:eval-e2e-lat}). Beyond this range, \sys scales to 1,024--8,192 devices---a regime where DTFM's solver exhausts memory and baseline communication architectures cannot amortize per-device overhead.
	We also evaluate the training time impact of device churn with \sys against SOTA edge (SWARM~\cite{swarm} and Asteroid~\cite{asteroid}) and cloud (Bamboo~\cite{bamboo} and Mario~\cite{mario}) methods.
	Our results show that, in our simulated single-failure settings, \sys achieves at least 100x faster recovery than prior methods.
\end{tightlist}

To summarize, \sys enables scalable and efficient training of large foundation models on edge devices by addressing the core challenges of memory constraints, communication overhead, and device heterogeneity, while delivering cloud-comparable per-batch runtime in our evaluated setting.

\section{Background and Motivation}
\label{sec:background}

\subsection{Edge Devices Characteristics}
\label{sec:background-device}
The edge can be broadly viewed as the set of devices outside datacenters~\cite{edgedl-review}.
For our purposes, we focus on the subset of edge devices that are network-connected, equipped with AI accelerators, and available while charging.
Examples include laptops and smartphones with integrated GPUs/NPUs (e.g., Apple M4/A16).
Such edge devices collectively offer compute capabilities comparable to the cloud, 
due to their massive scale~\cite{patterson-phone-cloud} (billions of units) and their substantial daily idle periods~\cite{phone-usage} (8~hrs on average).


\mypar{Memory limits}
Smartphone applications have strict memory limits.
Although vendors advertise 8--12GB of memory, each application can typically use at most 512MB~\cite{flexnn}.
Similarly, although laptops may advertise 32--64GB of memory, the usable memory available to training is often 10GB or less once fragmentation and background applications are accounted for.

\mypar{Device heterogeneity}
We consider the following:
(i)~\textit{Compute}, referring to variations in FLOPS performance across devices—for example, mobile devices offer 5–7 TFLOPS, while laptops like the Apple M3 Pro provide up to 27 TFLOPS~\cite{ai-benchmark}; and
(ii)~\textit{Communication}, which includes (1) differences in network link capacity, where both bandwidth and latency vary across devices~\cite{bt-broadband}, and (2) network link asymmetry, where per-device uplink (UL) speeds are typically 2–10× slower than downlink (DL), with typical downlink bandwidths of 10–100 MB/s and uplink bandwidths of 5–10 MB/s~\cite{speedtest}.

\mypar{Infrastructure-mediated coordination}
\review{
	Edge devices typically coordinate through infrastructure rather than via direct device-to-device links.
	They attach to access infrastructure and exchange training state through a managed relay, gateway, or cloud node.
	This matches recent edge~\cite{oakestra,luoshen} and RAN systems~\cite{edgeric,loca}, which place coordination at infrastructure components rather than on direct device-to-device links.
	Existing orchestration frameworks and cellular control infrastructure already follow this topology, with centralized coordination and shared state management.
}

\subsection{Training Characteristics}

\begin{table}[t]
	\centering
	\begin{tabular}{lcc}
		\hline
		\textbf{Model} & \textbf{GEMM TFLOPs} & \textbf{non-GEMM TFLOPs} \\
		\hline
		LLaMA 7B       & 5.613                & 0.038                    \\
		LLaMA 13B      & 9.768                & 0.048                    \\
		LLaMA 70B      & 27.096               & 0.083                    \\
		\hline
	\end{tabular}
	\caption{Floating-point operations for LLaMA models. Non-GEMM includes layer normalization, activation, and softmax.}
	\label{tab:llama_flops}
\end{table}

\mypar{GEMM-dominated training}
Across the LLaMA family, training compute is overwhelmingly concentrated in GEMMs.
As shown in Table~\ref{tab:llama_flops}, GEMMs account for more than 99\% of total training FLOPs for 7B, 13B, and 70B models, while layer normalization, activation functions, and softmax together contribute less than 1\%.
This establishes the first background fact behind \sys: any system that wants to accelerate training must primarily optimize GEMM execution.
\review{Each GEMM has an inherently asymmetric I/O profile: the two input matrices (rows of $\mathbf{A}$ and columns of $\mathbf{B}$) together require substantially more data movement than the compact partial output block---making each GEMM an \emph{input-heavy, output-light} computation.
}

\begin{table}[t]
	\centering
	\resizebox{\columnwidth}{!}{
		\begin{tabular}{|l|r|c|c|c|}
			\hline
			Stage               & Mem                                                                                        & Phone            & Laptop           & Cloud (A100)     \\
			                    &                                                                                            & (5\,TFLOPS)      & (27\,TFLOPS)     & (312\,TFLOPS)    \\
			\hline
			Fwd GEMM            & ${\approx}$76\,GB                                                                          & 3.9\,s           & 0.72\,s          & 0.063\,s         \\
			Fwd non-GEMM        & ${\approx}$26\,GB                                                                          & 26\,ms           & 9\,ms            & 0.7\,ms          \\
			Bwd GEMM            & ${\approx}$130\,GB                                                                         & 7.8\,s           & 1.44\,s          & 0.13\,s          \\
			Optimizer$^\dagger$ & \multicolumn{4}{c|}{Host-side, ${\approx}$2.25\,s (overlapped w/ Bwd, ${\approx}$338\,GB)}                                                          \\
			\hline
			\textbf{GEMM share} & \textbf{—}                                                                                 & \textbf{$>$99\%} & \textbf{$>$99\%} & \textbf{$>$99\%} \\
			\hline
		\end{tabular}
	}
	\captionof{table}{Per-step training time and memory reads for LLaMA-13B (batch 128, seq.\ 1024). Memory reads are hardware-independent; time varies by device TFLOPS. $^\dagger$Optimizer executes on the PS host~\cite{zero-offload}.}
	\label{tab:training-stages}
\end{table}

Table~\ref{tab:training-stages} breaks down one training step of LLaMA-13B, showing memory reads and compute time per stage across representative hardware.
Across all device classes, forward and backward GEMMs dominate per-step runtime by more than 99\%, while non-GEMM operators remain negligible in total compute time.
This stage-level view complements Table~\ref{tab:llama_flops}: the FLOP distribution generalizes across model sizes, and the runtime breakdown shows how that dominance appears in an actual training step.


\mypar{Operation placement by arithmetic intensity}
\review{
	GEMMs exhibit high arithmetic intensity (${\approx}$100--200\,FLOPs/byte for typical transformer tile sizes), making them compute-bound and well-suited to edge GPU/NPU acceleration.
	In contrast, optimizer updates (Adam), layer normalization, and softmax have low arithmetic intensity (${\approx}$1--2\,FLOPs/byte), making them memory-bandwidth-bound; the PS provides high-bandwidth host DRAM (DDR5, ${\approx}$150--200\,GB/s) and multi-core CPU capacity precisely matched to these requirements.
	Importantly, this is the same architectural choice made by cloud baselines: for Llama2-13B, the full training state (parameters + gradients + Adam optimizer states at 16\,bytes/param) totals ${\approx}$208\,GB, far exceeding A100 HBM capacity (40/80\,GB), so cloud frameworks such as ZeRO-Offload~\cite{zero-offload} also execute the optimizer on host CPU memory.
}



\subsection{Open Concerns}
\label{sec:memory-demand}

This subsection asks how existing parallelization schemes interact with edge memory budgets, asymmetric links, and device churn.
Distributed training commonly combines three standard forms of parallelism.
\textit{Data Parallelism (DP)} partitions input batches across full model replicas and synchronizes gradients via AllReduce~\cite{DBLP:conf/osdi/LiAPSAJLSS14,eddl};
\textit{Pipeline Parallelism (PP)} partitions the model by layers and pipelines micro-batches to reduce device idle time~\cite{gpipe,pipedream}; and
\textit{Tensor Parallelism (TP)} shards parameters within a layer across devices, reducing per-device memory demand while introducing additional collective communication~\cite{megatron,alpa}.

\begin{table}[t]
	\centering
	\begin{tabular}{|c|c|c|c|c|}
		\hline
		Model      & Total & Parameters & Optimizer & Activation \\
		\hline
		Llama2-7B  & 791GB & 12GB       & 48GB      & 731GB      \\
		Llama2-13B & 1.5TB & 24GB       & 95GB      & 1.4TB      \\
		Llama2-70B & 7TB   & 128GB      & 510GB     & 6.4TB      \\
		\hline
	\end{tabular}
	\captionof{table}{Total memory requirement for training, using batch size of 128, sequence length of 1024 and Megatron~\cite{megatron}.}
	\label{tab:memory-consumption}
\end{table}

\begin{table}[t]
	\centering
	\begin{tabular}{|c|c|c|c|c|}
		\hline
		Model      & DP    & PP    & DP+PP & DP+PP+TP        \\
		\#Devices  & 128   & 32    & 4K    & $>$8K           \\
		\hline
		Llama2-7B  & 65GB  & 25GB  & 1GB   & 50MB$\sim$500MB \\
		Llama2-13B & 128GB & 48GB  & 3GB   & 64MB$\sim$1GB   \\
		Llama2-70B & 688GB & 224GB & 14GB  & 140MB$\sim$5GB  \\
		\hline
	\end{tabular}
	\captionof{table}{Minimum per-device memory consumption for training using different parallelism modes, using~\autoref{tab:memory-consumption} setups. Typical usable memory limit of phone is 512MB~\cite{flexnn}.}
	\label{tab:training-memory-consumption}
\end{table}

\mypar{DP and PP still miss edge memory budgets}
\review{Foundation model training on the edge faces severe memory pressure.
	Table~\ref{tab:memory-consumption} shows that the full training state for LLaMA-13B totals 1.5\,TB, with activations alone accounting for 1.4\,TB and optimizer state adding a further 95\,GB.
	Table~\ref{tab:training-memory-consumption} then traces how parallelism strategies reduce per-device footprint: DP brings it to 128\,GB, PP to 48\,GB, and DP+PP to 3\,GB---still 6$\times$ the 512\,MB application memory limit of phones~\cite{flexnn}.
	Only when TP-class sharding is additionally applied does the per-device footprint drop to the 64\,MB--1\,GB range, spanning both phone and laptop budgets.
	Memory fitting is necessary, but only TP-class sharding achieves it.}

\begin{figure}[t]
	\includegraphics[width=.9\linewidth]{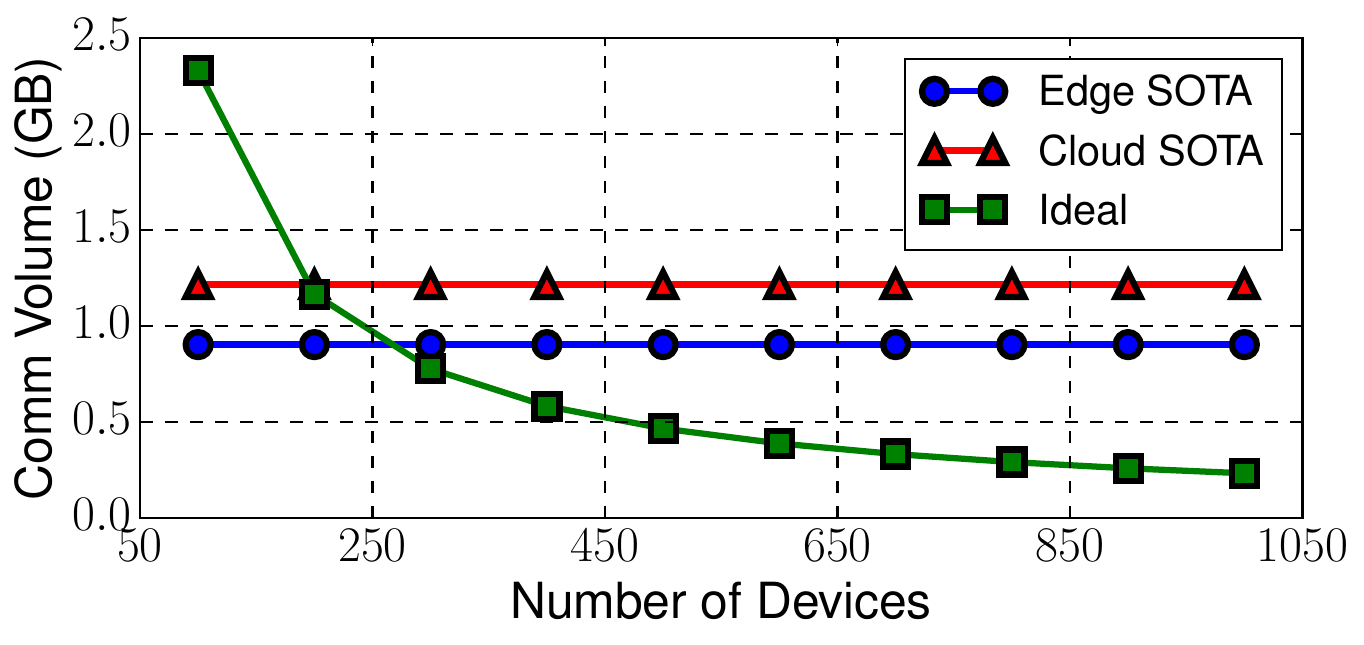}
	\caption{Per-device communication volume when training Llama2-13B with batch size 128 and sequence length 1024. The SOTA approaches are DTFM~\cite{dtfm} for the edge and Alpa~\cite{alpa} for the cloud. The additional communication volume of the cloud approach comes from AllReduce and AlltoAll at each layer.}
	\label{fig:comm-volume-problem}
\end{figure}


\mypar{Na\"{i}ve TP shifts the bottleneck to communication}
\review{Naively applying TP answers the memory question but creates a communication pattern poorly matched to the edge.
	As Figure~\ref{fig:comm-volume-problem} shows, standard TP introduces substantial per-device communication through collectives such as AllReduce and AlltoAll, causing communication demand to remain high even as more devices join---the opposite of the ideal scaling behavior.
	The challenge is therefore to preserve TP-like fine-grained sharding for memory efficiency while ensuring per-device communication volume \emph{decreases} with scale.}

\mypar{Static placement struggles with heterogeneity and churn}
Edge training pools are inherently heterogeneous, spanning a 5.4$\times$ compute range in our setting (5--27\,TFLOPS; \S\ref{sec:background-device}).
In synchronous TP, equal sharding makes step time depend on the slowest participant, so this capability gap directly translates into idle time for faster devices.
For example, when equal work is assigned across such devices, a phone-class participant can take about 5$\times$ longer than a laptop-class participant on the same shard, leaving the faster devices idle for most of the step.

The problem is compounded by churn.
Even with a conservative interruption rate of 1\% per device per hour, the system-level mean time between failures drops to about 47 minutes at 128 devices, 12 minutes at 512 devices, and below 6 minutes at 1{,}024 devices.
Standard collectives cannot complete when a participant disappears, so restarting the full step or relying on heavyweight checkpoint recovery wastes partial work and quickly becomes impractical at scale.
\subsection{Existing Solutions and their Limitations}

Table~\ref{tab:related-comparison} summarizes how prior systems align with the requirements of edge FM training.
Across these families, existing approaches do not simultaneously satisfy edge memory budgets, communication scalability, and robustness to heterogeneity and churn.

\mypar{Handling parallelism}
Existing distributed training approaches fall into two broad categories: native cloud approaches and cloud approaches adapted to the edge.
Current cloud-based systems (\eg \cite{megatron,alpa,atp}) combine DP, PP, and TP.
This combination can reduce per-device memory demand, but it assumes homogeneous compute and communication resources together with high-bandwidth interconnects such as NVLINK.
These assumptions do not hold in the edge environment, so applying cloud systems directly leads to poor performance.
Existing edge training systems therefore retreat to 2D parallelism, specifically DP~\cite{eddl,dedloc} and PP~\cite{dtfm,swarm}.
They avoid TP because peer-to-peer TP communication overwhelms its memory benefit on constrained links, as shown in~\autoref{fig:comm-volume-problem}.
Consequently, neither family resolves the full edge setting: cloud 3D methods depend on datacenter-style networking, while edge DP/PP methods retain high per-device memory demand and DP AllReduce overhead.

\mypar{Handling heterogeneity}
With regard to device heterogeneity, existing training systems address different aspects.
AMP\cite{amp}, HetPipe\cite{hetpipe}, and SDPipe\cite{sdpipe} focus on compute heterogeneity using PP.
DTFM\cite{dtfm} and FusionAI\cite{fusionai} target network heterogeneity, with DTFM combining DP and PP.
DeDLOC\cite{dedloc} optimizes DP AllReduce for communication heterogeneity but uses asynchronous updates, risking accuracy loss.

\mypar{Handling device churn}
Existing approaches for handling device churn either employ checkpoint-restore, or replication and recomputation.
Mario~\cite{mario} and Tenplex~\cite{tenplex} represent the cloud methods of checkpointing all training states.
Bamboo~\cite{bamboo} replicates the per-layer compute in the context of spot instances.
SWARM~\cite{swarm} adopts rewiring, where failed hidden states are rerouted to devices holding the same layer for recomputation. Asteroid~\cite{asteroid} employs resharding and redistributed layers in addition to recomputation.
Learning@home~\cite{learning@home} has a particular focus on expert recompute for Mixture-of-Expert type foundation models.

\mypar{Alternative training paradigms}
Federated Learning (FL)~\cite{fedscale,photon} is akin to DP and has a focus on data privacy, making it orthogonal to FM training on edge devices with public data leveraging additional forms of parallelism. Our focus is on the latter.
The fault tolerance strategy in FL can be lossy due to dropping gradients, similar to asynchronous gradients~\cite{moshpit}, while our focus is on the fully synchronous and lossless version.
While incorporating Split Learning (SL) adds scalability and capacity with offloading to the cloud~\cite{splitfed}, it is still equivalent to the DP+PP scheme, falling short in performance and accuracy.

Recent work on distributed LLM training takes a complementary local-SGD approach. DiLoCo~\cite{diloco} and OpenDiLoCo~\cite{opendiloco} let each device run many local steps on its shard before syncing through a lightweight outer optimizer such as Nesterov momentum. This removes per-step communication, but it also introduces staleness between rounds. \sys and local SGD therefore sit at different points in the synchrony--communication trade-off: \sys keeps fully synchronous training, with identical gradients at every step, while DiLoCo relaxes synchrony to cut communication. The two are complementary; a hybrid that combines \sys's fine-grained GEMM sharding with periodic synchronization from DiLoCo is an interesting direction.

Taken together, prior systems do not simultaneously fit edge memory budgets, reduce per-device communication with scale, and recover efficiently under heterogeneity and churn. This gap motivates \sys's design choice: combine PS-mediated coordination with fine-grained sub-GEMM scheduling so that memory efficiency, communication efficiency, and recovery all follow from the same execution abstraction.

\begin{table*}[t]
	\centering
	\caption{Approaches for edge FM training against the requirements
		and challenges}.
	\label{tab:related-comparison}
	\resizebox{\linewidth}{!}{%
		\begin{tabular}{l c c c c c c}
			\toprule
			\textbf{Property}
			                             & \makecell{\textbf{Cloud 3D}                                                                              \\ {\cite{megatron,alpa,atp}}}
			                             & \makecell{\textbf{Edge DP/PP}                                                                            \\ {\cite{dtfm,eddl,amp,hetpipe,dedloc,swarm}}}
			                             & \makecell{\textbf{Ckpt/Repl.}                                                                            \\ \textbf{FT} \\ {\cite{mario,bamboo,asteroid,tenplex}}}
			                             & \makecell{\textbf{Federated}                                                                             \\ \textbf{Learning} \\ {\cite{fedscale,photon}}}
			                             & \makecell{\textbf{Local-SGD}                                                                             \\ {\cite{diloco,opendiloco}}}
			                             & \makecell{\textbf{\sys}                                                                                  \\ {(this work)}} \\
			\midrule
			Per-device mem.\ fits edge   & needs NVLINK                  & layer-bound    & replica cost & full replica    & full replica  & \cmark \\
			Comm.\ scales w/ \#devices   & \xmark                        & \xmark         & \xmark       & \xmark          & periodic only & \cmark \\
			UL/DL asymmetry-aware        & \xmark                        & \xmark         & \xmark       & \xmark          & \xmark        & \cmark \\
			Exact gradient semantics     & \cmark                        & async (DeDLOC) & \cmark       & lossy agg.      & stale grads   & \cmark \\
			Compute heterogeneity        & \xmark                        & PP-level       & \xmark       & drop stragglers & \xmark        & \cmark \\
			Network heterogeneity        & \xmark                        & DeDLOC only    & \xmark       & \xmark          & \xmark        & \cmark \\
			Sub-layer fault recovery     & \xmark                        & full-layer     & full ckpt    & drop grads      & lose local    & \cmark \\
			Scales beyond batch/\#layers & cloud HW only                 & \xmark         & \xmark       & \xmark          & \cmark        & \cmark \\
			\bottomrule
		\end{tabular}%
	}
\end{table*}

\section{\sys Training Framework and Method}
\label{sec:sys_overview}

\subsection{Key Insights}
\label{sec:obervations}

\mypar{GEMM I/O aligns with edge asymmetry}
\review{The opportunity comes from what GEMM dominance \emph{enables}: each GEMM has a structurally asymmetric I/O pattern. The two input matrices---rows of $\mathbf{A}$ and columns of $\mathbf{B}$ assigned to a device---are substantially larger than the compact partial output block returned by that device, as confirmed by Table~\ref{tab:llama_gemm_shapes}. In a parameter-server-centric deployment, the PS dispatches inputs over downlink and collects outputs over uplink, directly matching the link asymmetry of edge networks where downlink bandwidth exceeds uplink by 2--10$\times$ (\S\ref{sec:background-device}). This \emph{structural alignment} means that as more devices join, each device handles fewer rows and columns per GEMM, so per-device communication volume \emph{decreases} toward the ideal scaling behavior in Figure~\ref{fig:comm-volume-problem}, which conventional TP collectives (AllReduce, AlltoAll) cannot achieve.}

\mypar{Sub-GEMM scheduling unifies the design}
\review{The same decomposition also provides a unified scheduling abstraction. Since GEMMs within a DAG level are mutually independent (Table~\ref{tab:llama_gemm_shapes}), training can be decomposed into independent row/column sub-tasks. This single abstraction simultaneously addresses (a)~\emph{memory}---each device holds only its assigned row/column shards, reducing per-device footprint to the range achievable by TP (\S\ref{sec:background}); (b)~\emph{communication}---total GEMM volume is bounded per operation, so the per-device share shrinks as the device count grows; and (c)~\emph{fault tolerance}---a failed device loses only its assigned shards, which are redistributed via the same cost model used for normal scheduling, enabling fast fine-grained recovery (\S\ref{sec:device-churn}). \emph{Thus, \sys should schedule training at sub-GEMM granularity: shard each GEMM into row/column tasks and let the PS dispatch inputs, collect partial outputs, and recover lost shards through the same scheduler.}}

\begin{table}[t]
	\centering
	\begin{tabular}{|lcccc|}
		\hline
		\textbf{Component}       & \textbf{M} & \textbf{K} & \textbf{N} & \textbf{Count}  \\
		\hline
		Q/K/V projection         & 1024       & 4096       & 4096       & 128 $\times$ 3  \\
		Q × K\textsuperscript{T} & 1024       & 128        & 1024       & 128 $\times$ 32 \\
		\hline
		MLP up-proj              & 1024       & 4096       & 11008      & 128             \\
		\hline
	\end{tabular}
	\caption{Representative GEMMs and counts in a single transformer layer during forward propagation; the same observation applies to backward propagation. We use batch size 128 and sequence length 1024. There are no memory dependencies among GEMMs with the same $(M,K) \times (K,N)$ shape.}
	\label{tab:llama_gemm_shapes}
\end{table}

This idealized method represents training as a series of operators in a directed acyclic graph (e.g., as in Alpa), with a controller distributing operator execution across devices. The communication load then consists of model parameters and all intermediate results. Because devices return the output of each operator directly to the controller, gradients can be aggregated there without peer-to-peer broadcasts.
In addition, each parameter gradient and each layer's intermediate result is transmitted only once, so the total communication per batch becomes model size + (intermediate size $\times$ number of layers).


\subsection{\sys Design}
\label{sec:overview}

\sys approximates the idealized method outlined above using a client-server communication architecture, rather than the peer-to-peer communication adopted by cloud-based and current edge distributed-training approaches.
At the core of \sys is a parameter server (PS) that, as a server, plays several key roles: (i) scheduling training operations across devices, (ii) maintaining training parameters and eliminating aggregation over network (AlltoAll and AllReduce) operations, and (iii) tracking device availability and capabilities (compute resources and downlink/uplink speeds).
Devices as clients carry out training operations assigned by the PS, and tensor parallelism enables an arbitrary number of participants.

We term this strategy \emph{selective hybrid tensor parallelism}, where ``selective'' denotes using the same cost model for both heterogeneity-aware scheduling and churn recovery (\S\ref{sec:gemm_scheduling}, \S\ref{sec:device-churn}).
When assigning even a single row--column pair to a straggler would degrade overall performance, \sys redistributes that work to more capable devices to maintain system efficiency.
Likewise, if a device fails mid-round, the unfinished sub-GEMMs are redistributed across the remaining live devices through the same scheduling machinery, while newly joined devices enter on the next GEMM round.

\begin{figure}[t]
	\centering
	\includegraphics[width=.92\linewidth]{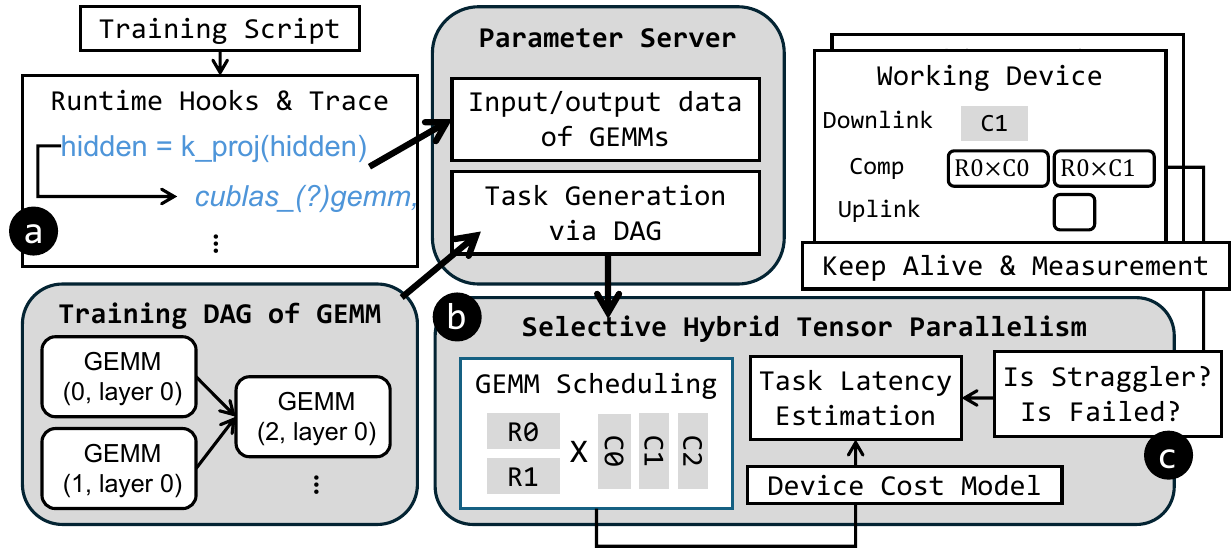}
	\Description{Diagram showing CLEAVE's workflow: a training script is traced into a DAG of GEMM operations, which are then selectively scheduled across heterogeneous edge devices with overlapped communication and computation.}
	\caption{Workflow of \sys, from a model defined in a training script to a DAG of GEMMs. Edges in the DAG represent memory dependencies. Each GEMM is scheduled selectively across devices while overlapping communication and computation where possible. The workflow is traced from runtime GEMM calls and then mapped to selective hybrid tensor-parallel scheduling across heterogeneous devices.
	}
	\label{fig:transformer-dag}
\end{figure}

\mypar{Seamless training as in cloud}
\sys first attaches runtime hooks to the training script to trace GEMM calls (e.g., \texttt{cublas\_gemm} from linear layers), yielding a DAG where nodes are GEMMs and edges encode data dependencies~\cite{alpa,amp,fusionai} (in~\autoref{fig:transformer-dag}~\myc{a}).
This DAG can be extracted using software hooks (e.g., PyTorch) to trigger custom scheduling routines.
For example, the QKV projections in self-attention are three GEMMs with no inter-dependencies, while the outputs of the Q and K projections feed another GEMM.
\sys preserves the size and numerical semantics of each GEMM to ensure consistency with standard cloud-based training.
Because no compression or gradient approximation is applied, the computed gradients are mathematically equivalent to those produced by a single-device execution; in practice, floating-point rounding differences across heterogeneous accelerators may introduce bit-level variation consistent with standard non-determinism in distributed training~\cite{megatron}. Consequently, \sys preserves the same update semantics as cloud-based training for any given model, optimizer, and hyperparameter configuration, subject to the usual nondeterminism of distributed floating-point execution.
The PS then performs task generation over this DAG: for each GEMM node, it creates row and column blocks from the data captured by the GEMM calls with zero copy.
At each DAG node, the PS reads pre-computed decisions from the cost model to determine optimal scheduling, dispatches the corresponding rows and columns to participating devices, and aggregates the computed outputs.
When the device set changes, \sys overlaps cost-model recomputation with device-side computation and updates the scheduling decisions for the next batch.

\mypar{Scheduling Workflow}
\sys requires devices to register upon joining and report their compute and communication capabilities, including uplink (UL) and downlink (DL) bandwidth.
Since GEMM shapes repeat across layers, the cost model optimization is solved once per device set and reused thereafter.
Our hybrid tensor parallelism adapts DL-to-UL bandwidth asymmetry for each device by shaping the row–column assignments accordingly (in~\autoref{fig:transformer-dag}~\myc{b}).
To maximize efficiency, \sys overlaps communication and computation using parallel threads and a stream-based protocol (\eg gRPC~\cite{grpc}): rows and columns are downloaded, GEMMs computed, and results uploaded concurrently.
Device departures are detected via disconnect events; unfinished GEMM outputs are identified using bookkeeping, and the remaining work is rescheduled by re-invoking the cost model.
Non-GEMM operations (layer normalization, activation functions, softmax) are computed locally on the PS, with negligible compute overhead (Table~\ref{tab:llama_flops}).
This placement is deliberate: \sys offloads compute-bound GEMMs to edge accelerators while keeping memory-bandwidth-bound stages such as optimizer updates and other non-GEMM operators on the PS's host DRAM and CPU~\cite{zero-offload}.

\mypar{Heterogeneity-aware scheduling and churn recovery}
\sys treats heterogeneity-aware scheduling and churn recovery as two views of the same scheduling problem over the GEMM DAG (in~\autoref{fig:transformer-dag}~\myc{c}).
When a device joins, it registers its compute and bandwidth; the PS incorporates this into the cost model and, on the next optimization run, skews work toward stronger devices.
During training, devices periodically send keep-alive and measurement signals; if a device fails mid-batch, the scheduler identifies the unfinished sub-GEMMs and re-solves a smaller optimization problem to redistribute the remaining row--column blocks across the live devices.
Because GEMMs are sharded at a fine granularity, only a small fraction of a GEMM in one layer must be recomputed on a single point of failure, enabling fast recovery.
\review{This fine-grained recovery scope is a deliberate design choice: by making the unit of work a sub-GEMM shard rather than a full model layer, \sys bounds the blast radius of each device failure to a small fragment that can be redistributed across all remaining devices.}
The full churn-recovery formulation and cache-aware communication cost are detailed in~\S\ref{sec:device-churn}.

\section{\sys Scheduling Methodology}
\label{sec:cost-model}

\subsection{Optimization Problem and its Cost Model}
\label{sec:gemm_scheduling}
In this section, we formalize how \sys schedules the traced GEMM DAG over a heterogeneous device set. Given the participating devices $\mathbf{D}$ and the GEMMs extracted in~\S\ref{sec:overview}, the scheduler chooses how much row/column work of each GEMM to assign to each device so as to minimize the completion time of the full training batch. We first define the level-wise objective, then specify the communication and computation terms, and finally state the feasibility constraints. The same formulation is later reused for churn recovery on a smaller subproblem.

\mypar{Problem definition and objective}
We process the DAG of GEMMs in \emph{level} order, where GEMMs at the same level $s$ have the same critical-path distance from the root of the DAG, which is the start of a training batch. For instance, when the model enters the first layer (shown in~\autoref{fig:transformer-dag}), the QKV projections ($XW_Q$, $XW_K$, $XW_V$) lie at the same level. GEMMs within a level have no memory dependency and can therefore execute in parallel. The output of level $s-1$ becomes the input of level $s$, so level $s+1$ cannot start before level $s$ finishes. The optimization thus minimizes the completion time of the final level by recursively composing per-level costs.

We define $C_{\text{GEMM}}(s)$ as the latency from level $0$ to level $s$. Within level $s$, multiple GEMMs can run in parallel, and the cost of GEMM $p$ in that level is denoted by $C_{\text{GEMM}}(s,p)$. For the first level, the latency is simply the maximum GEMM cost within that level. For each later level, the latency accumulates recursively from the predecessor level:
\begin{align}
	C_{\text{GEMM}}(0)   & := \max\limits_{p} C_{\text{GEMM}}(0,p), \nonumber                                   \\
	C_{\text{GEMM}}(s)   & := C_{\text{GEMM}}(s-1) + \max\limits_{p} C_{\text{GEMM}}(s,p), \quad s>0, \nonumber \\
	C_{\text{GEMM}}(s,p) & := \max\limits_k \{ C_{\text{GEMM}}(s,p,k) \} \label{eq:level-cost-s}
\end{align}
We therefore treat the slowest GEMM in a level as the latency of that level. \review{The scheduler minimizes the distributed GEMM completion term $C_{\textsc{GEMM}}(S-1)$, but the full end-to-end batch time additionally includes the exposed PS-side optimizer tail $C_{\textsc{OptTail}}^{\textsc{PS}}$, defined below. Accordingly, we write the batch time as $C_{\textsc{Batch}} := C_{\textsc{GEMM}}(S-1) + C_{\textsc{OptTail}}^{\textsc{PS}}$.}

\mypar{Search space of GEMM}
For each GEMM in level $s$, the PS decides how many rows and columns to send to each device so that the cost above is minimized. Formally, each GEMM in level $s$ has two inputs $\mathbf{A}_s \in \R^{m_s\times n_s}, \mathbf{B}_s \in \R^{n_s\times q_s}$ and an output $O_s \in \R^{m_s\times q_s}$. The matrix elements use the same numerical precision (\eg FP16 or INT8) with byte size $b$ (\eg FP16 has $b=2$). We split $\mathbf{A}_s$ by rows (DP-style sharding) and $\mathbf{B}_s$ by columns (TP-style sharding). Each device $d_k$ computes $\mathbf{A}_{s,p,k}'\circ \mathbf{B}_{s,p,k}' \subset O_s$, with $\mathbf{A}_{s,p,k}'\in \R^{\alpha_{s,p,k}\times n_s}$ and $\mathbf{B}_{s,p,k}'\in \R^{n_s\times \beta_{s,p,k}}$. Collectively, $\cup_k \mathbf{A}_{s,p,k}'\circ \mathbf{B}_{s,p,k}' \equiv O_s$.

\mypar{Communication cost}
The UL and DL communications can be overlapped with one another as well as with computation, as captured in~\autoref{eq:communication-cost}.
\begin{align}
	C_{\text{GEMM}}(s,p,k) := \max\{ & C^d_{\textsc{Comm}}(s,p,k), \nonumber \\ & C^u_{\textsc{Comm}}(s,p,k), C_{\textsc{Comp}}(s,p, k)\}\label{eq:communication-cost}
\end{align}
The communication terms capture two per-device properties: (i)~\emph{overhead}, including network-protocol and end-to-end transfer delay, and (ii)~\emph{bandwidth}, covering both UL (sending data to the PS) and DL (receiving data from the PS). We model these quantities as device-associated constants, assuming the device is charging at a fixed site with a stable network connection. Suppose $W^u \in \R_{+}^{|\mathbf{D}|}$ and $W^d \in \R_{+}^{|\mathbf{D}|}$ are the UL and DL bandwidth vectors for all devices. Similarly, $L^u \in \R_{+}^{|\mathbf{D}|}$ and $L^d \in \R_{+}^{|\mathbf{D}|}$ are the UL and DL latency-overhead vectors. For each device $d_k \in \mathbf{D}$, the communication cost is:
\begin{align}
	C_{\textsc{Comm}}^d(s, p, k) & := \frac{\alpha_{s,p,k}n_sb}{W^d_k}  + \frac{n_s\beta_{s,p,k}b}{W^d_k}  + L^d_k, \nonumber \\
	C_{\textsc{Comm}}^u(s, p, k) & := \frac{\alpha_{s,p,k}\beta_{s,p,k}b}{W^u_k} + L^u_k
	\label{eq:ul-dl-cost}
\end{align}
Here, $W^u_k \in W^u$ and $W^d_k \in W^d$ denote the UL and DL bandwidth of device $d_k$, and $L^u_k \in L^u$ and $L^d_k \in L^d$ are the corresponding UL and DL overhead terms. $b$ is the byte size of the matrix elements; \eg BF16 means two bytes per element.

These terms map directly to the training pipeline. During the \emph{forward pass}, the PS dispatches each device's assigned row/column slices ($\mathbf{A}_s$ and $\mathbf{B}_s$ shards) over downlink. During the \emph{backward pass}, each device uploads partial outputs over uplink for gradient collection at the PS. This accounting captures weight distribution and partial-output aggregation without requiring peer-to-peer collectives.

\mypar{Computation cost}
The computation term captures the on-device GEMM time for each assigned shard:
\begin{align}
	C_{\textsc{Comp}}(s,p, k) := \frac{2}{F_k}\alpha_{s,p,k}\beta_{s,p,k}n_s \label{eq:computation-cost}
\end{align}
For each device $d_k \in \mathbf{D}$, with FLOPS capability $F_k$, this term is proportional to the FLOPS required for the assigned sub-GEMM and uses the standard $2mnq$ GEMM count~\cite{scaling-law}. \review{We keep non-GEMM operators and optimizer updates on the PS, as described in~\S\ref{sec:overview}. Although they do not affect the shard-placement decision itself, we model their exposed coordinator-side cost explicitly rather than assuming it away. For a GEMM weight matrix $\mathbf{B}_s \in \R^{n_s\times q_s}$, let $\rho_{\textsc{Opt}}$ denote the host-memory traffic per parameter update (26 bytes/parameter for Adam with BF16 weights, gradients, and first/second moments in our evaluation), and let $B_{\textsc{PS}}^{\textsc{mem}}$ denote the effective PS host-memory bandwidth. The corresponding PS-side optimizer time is}
\begin{align}
	C_{\textsc{Opt}}^{\textsc{PS}}(s,p) := \frac{\rho_{\textsc{Opt}} n_s q_s}{B_{\textsc{PS}}^{\textsc{mem}}}. \label{eq:optimizer-cost}
\end{align}
\review{Because \sys pipelines optimizer work by DAG level behind backward GEMM execution, only the final unhidden stage contributes to end-to-end batch time. We therefore model the exposed optimizer contribution as}
\begin{align}
	C_{\textsc{OptTail}}^{\textsc{PS}} := \max\limits_{s,p} C_{\textsc{Opt}}^{\textsc{PS}}(s,p). \nonumber
\end{align}
\review{This keeps the placement objective focused on the distributed GEMM schedule while making the PS-side optimizer cost explicit in the end-to-end batch-time model.}

\mypar{Cost model constraints}
The shard assignments across all devices must cover the full GEMM output without redundant work:
\begin{align}
	\sum_{k=1}^{N}\alpha_{s,p,k}\beta_{s,p,k} = m_sq_s. \nonumber
\end{align}
Furthermore, a straggler may be excluded entirely by allowing a device to stay idle instead of forcing it to process an unhelpfully small shard. We express this as:
\begin{align}
	\st \; & (\alpha_{s,p,k} = 0 \land \beta_{s,p,k} = 0) \lor (\alpha_{s,p,k} \neq 0 \land \beta_{s,p,k} \neq 0) \label{eq:zero-task-constraint}
\end{align}
This covers cases where reducing load on a weak device is still insufficient and the scheduler should leave that device idle.

The cost model is also constrained by device memory capacity. Let $\mathbf{M} \in \R_{+}^{N}$ be the memory capacity of devices in bytes. Because all dispatched rows and columns must remain on device until the corresponding pairwise products are computed, we enforce:
\begin{align}
	\st \; & \alpha_{s,p,k}n_sb + n_s\beta_{s,p,k}b + \alpha_{s,p,k}\beta_{s,p,k}b \le M_k \label{eq:memory-constraint}
\end{align}
Here, $M_k \in \mathbf{M}$ is the memory capacity of device $d_k$. We obtain the scheduling decision using Gurobi~\cite{gurobi}.

\mypar{Solver and reuse}
\review{
The solver operates in two regimes, summarized in Table~\ref{tab:milp-churn}. In the \emph{cold-start phase}, \sys solves the full optimization once for a device set and reuses the result across repeated GEMM shapes. In the \emph{online phase}, churn handling solves a much smaller incremental subproblem that reassigns only orphaned shards while keeping surviving assignments fixed. For the largest configuration we evaluate (1,024 devices and a 70B-parameter model), the cold-start solve takes approximately 10 minutes and is amortized across thousands of training batches, while churn-time re-optimization completes in seconds.}

\begin{table}[t]
	\centering
	\caption{Initial cold-start optimization versus churn-time incremental re-optimization.}
	\label{tab:milp-churn}
	\resizebox{\columnwidth}{!}{
		\begin{tabular}{lcc}
			\toprule
			                   & \textbf{Initial cold-start}              & \textbf{Churn re-solve (1 device)}               \\
			\midrule
			Devices considered & 1,024                                    & $\sim$1,023 (fixed assignments)                  \\
			Shards to assign   & Thousands--millions                      & Dozens                                           \\
			Decision variables & $O(\text{devices} \times \text{shards})$ & $O(\text{devices} \times \text{failed\_shards})$ \\
			Solve time         & $\sim$10 min                             & Seconds                                          \\
			\bottomrule
		\end{tabular}
	}
\end{table}

\subsection{Churn recovery}
\label{sec:device-churn}
\sys reuses the cost model from~\autoref{sec:gemm_scheduling} to solve the churn-recovery problem that arises when devices fail mid-batch. We use the same solver as in~\autoref{sec:gemm_scheduling}, but on a smaller instance induced by the unfinished shards.


We treat each device-failure event as a new snapshot of the scheduling problem.
Suppose that a failed device was responsible for computing a subset of the GEMM, denoted by $\mathbf{A}_s^* \subseteq \mathbf{A}_s'$ and $\mathbf{B}_s^* \subseteq \mathbf{B}_s'$.
Our goal is to reschedule the computation of the submatrix $\mathbf{A}_s^* \circ \mathbf{B}_s^*$.
We define binary matrices $\mathbf{R}_s \in \{0,1\}^{|\mathbf{D}| \times m_s}$ and $\mathbf{C}_s \in \{0,1\}^{|\mathbf{D}| \times q_s}$ to represent the presence of row and column caches on each device, corresponding to mappings of $\mathbf{A}_s$ and $\mathbf{B}_s$ across the device set $\mathbf{D}$.
That is, $\mathbf{R}_s[k, i] = 1$ if device $k$ holds row $i$ of $\mathbf{A}_s$, and similarly $\mathbf{C}_s$ reflects cached columns of $\mathbf{B}_s$.
Given the memory constraint in~\autoref{eq:memory-constraint}, we assume that no cache replacement occurs during the execution of a GEMM at level $s$. Accordingly, we enforce the constraints:
$\alpha_{s,k} = \sum_i \mathbf{R}_s[k, i], \quad \beta_{s,k} = \sum_j \mathbf{C}_s[k, j]$,
where $\alpha_{s,k}$ and $\beta_{s,k}$ denote the number of row and column blocks cached on device $k$, respectively.

We aim to handle both single-device and multiple simultaneous device failures.
For all failed row and column computations across devices, we concatenate the affected blocks to form $\mathbf{A}_s^*$ and $\mathbf{B}_s^*$. This reduces the problem to the same scheduling formulation described in~\autoref{sec:gemm_scheduling}, but with a cache-aware communication term for the unfinished work.
Let $\hat{\alpha}_{s,p,k}$ and $\hat{\beta}_{s,p,k}$ denote the numbers of row and column blocks that device $k$ still needs to fetch for the failed shard after accounting for its local caches. For each device $d_k \in \mathbf{D}$, the DL communication cost becomes:
\begin{align*}
	C_{\textsc{Comm}}^d(s, p, k) := \frac{\hat{\alpha}_{s,p,k}n_sb}{W^d_k} + \frac{n_s\hat{\beta}_{s,p,k}b}{W^d_k}
\end{align*}

Here, $n_s$ is the size of the current GEMM, $b$ is the block size, and $W^d_k$ is the downlink bandwidth of device $d_k$. If a required row or column is already cached on the target device, the corresponding term is zero; otherwise, only the missing blocks contribute to the transfer time. The uplink cost $C_{\textsc{Comm}}^u(s, p, k)$ is computed analogously for the partial output.


\section{Evaluation}
\label{sec:evaluation}

\subsection{Evaluation Settings}
We evaluate \sys through simulation of large-scale scenarios with high device heterogeneity.
Simulation-based evaluation is standard for edge-training systems (\eg DTFM~\cite{dtfm}, FusionAI~\cite{fusionai}) because deploying thousands of heterogeneous devices is infeasible in practice.
Our evaluation focuses on the GEMM DAG traced from the HuggingFace Trainer by applying hooks to linear layers and matrix multiplication in the considered models.
We use OPT~\cite{opt} and Llama2~\cite{llama2} models of various sizes.

Our primary metric is \textbf{per-batch runtime of the distributed GEMM pipeline}.
It includes device-side forward and backward GEMM computation, communication, and scheduling overhead.
\review{We exclude device-side non-GEMM layers (\eg LayerNorm and activation functions) from this GEMM-centric runtime accounting because their on-device computation and communication cost is negligible relative to GEMMs ({<}1\% of FLOPs, Table~\ref{tab:llama_flops}). Different systems handle them differently (\eg local in DTFM versus distributed in Alpa), which would confound comparisons if included. However, we do \emph{not} ignore coordinator-side optimizer work: we explicitly model the exposed PS-side optimizer tail $C_{\textsc{OptTail}}^{\textsc{PS}}$ in \autoref{sec:cost-model} and account for it separately below.}

We choose the following baselines:
(i)~\textbf{DTFM}, representing edge training with heterogeneity-aware DP and PP;
(ii)~\textbf{Alpa}, representing cloud training using DP, PP and TP, assuming homogeneous devices.
For a given set of devices, all baselines and \sys work out a scheduling plan for traversing the GEMM DAG, \ie the volume of data sent from each device and the computation FLOPS assigned to it.
Because published baseline cost models do not directly account for both network and computation latency, we evaluate all methods under the same latency accounting model.


If not specifically mentioned, training is set to a batch size of 128 and sequence length of 1024, based on common settings~\cite{opt}.
For \sys, we assume the PS has data center capability with 200 Gbps network bandwidth and CPUs with 128 cores.
Device network and compute capabilities are sampled from the datasets described in \S\ref{sec:background-device}.
All results reported are the average of multiple simulation runs.


\subsection{Training Performance}

\begin{figure}[t]
	\centering
	\centering
	\includegraphics[width=\linewidth]{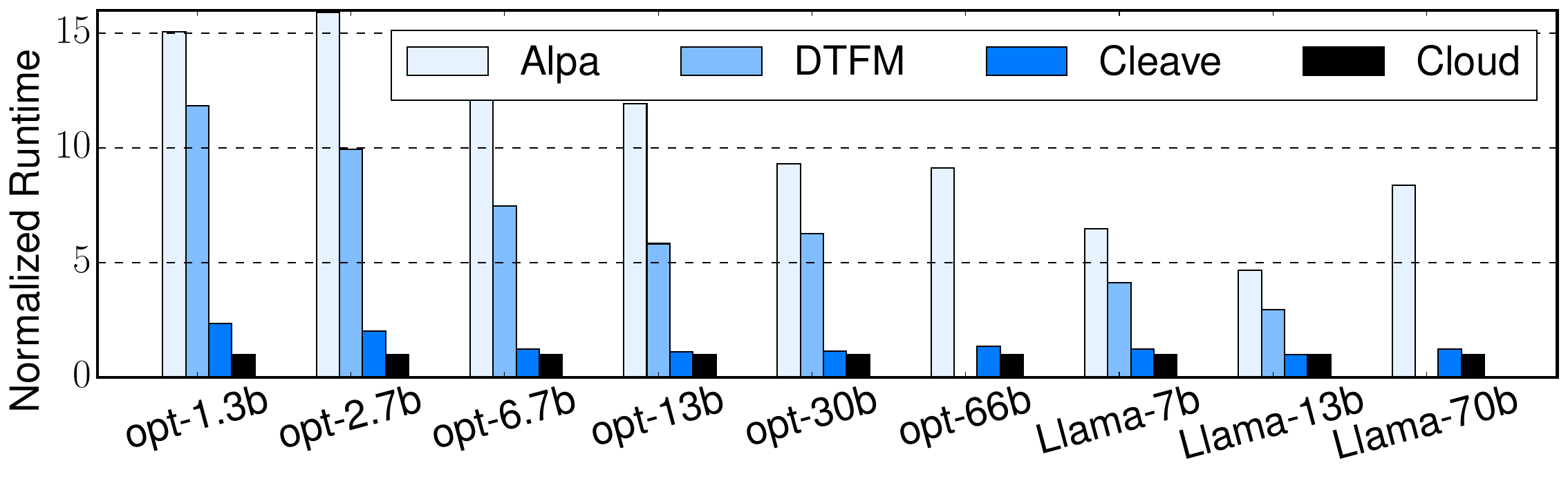}
	\caption{Normalized per-batch runtime (lower is better). Under the matched-resource methodology of \S\ref{sec:evaluation}, \sys achieves cloud-comparable performance while the baselines do not.
	}
	\label{fig:eval-e2e-lat}
\end{figure}

Using the per-batch runtime metric defined above, we ask whether \sys and the baselines can achieve cloud-comparable performance under matched resource envelopes while supporting large models at a fixed number of devices.
\review{As defined above, the plotted per-batch runtime includes device-side forward and backward GEMM computation; PS-to-device communication during the forward pass (weight dispatch) and device-to-PS communication during the backward pass (output shard upload)---both as defined in the cost model (\autoref{sec:cost-model}); and scheduling overhead. In addition, we explicitly model the coordinator-side optimizer term $C_{\textsc{OptTail}}^{\textsc{PS}}$ from \autoref{sec:cost-model} rather than assuming the PS-side optimizer cost is zero.
	For GEMM-offloaded training in \sys, Adam updates run on the PS host CPU and are pipelined by DAG level behind backward GEMM execution. For Llama2-13B, the monolithic host-side Adam traffic is ${\sim}$338\,GB per step, corresponding to ${\sim}$2.25\,s at 150\,GB/s host-memory bandwidth; layer-wise pipelining reduces the exposed tail to approximately 56\,ms, which is less than 0.1\% of a 60--120\,s batch. This is not a differential penalty against the single-GPU cloud baseline because DeepSpeed ZeRO-Offload also executes optimizer updates in host memory~\cite{zero-offload}.}
To ensure a fair comparison across heterogeneous resource envelopes, we apply two standard normalizations before runtime comparison: (i) we align the aggregate network bandwidth of edge devices with that of cloud GPUs---matching PCIe bandwidth when the model does not fit entirely in GPU memory, or InfiniBand/NVLink bandwidth otherwise; and (ii) we align average achieved FLOPS, since individual device utilization is lower under static workload partitioning even though collective FLOPS may be high. This isolates scheduling and communication efficiency---\sys's core contribution---rather than conflating it with raw hardware provisioning differences. As a robustness check beyond this controlled setting, Appendix~B scales edge devices proportionally with cloud GPU count \emph{without} alignment and confirms that \sys scales effectively even outside this controlled environment.
\noindent\textbf{Example.}
For Llama2-13B trained with 512 edge devices ($F_k\approx6$\,TFLOPS, $W^d_k\approx55$\,MB/s, $W^u_k\approx7.5$\,MB/s), the aggregate achieved edge FLOPS is $512\times6\times\eta_{\mathrm{edge}}\approx512\times1.8=922$\,TFLOPS (using typical 30\% utilization); this is aligned to $3\times$A100 at $312$\,TFLOPS each. Aggregate DL bandwidth is $512\times55\,\text{MB/s}\approx28$\,GB/s, aligned to PCIe\,4.0 at 32\,GB/s. Under these aligned parameters, the normalized per-batch time reported in Figure~\ref{fig:eval-e2e-lat} represents scheduling and communication efficiency with matched total hardware capacity.

For the cloud-based alternative, we use DeepSpeed~\cite{deepspeed} as the training framework, which supports virtually unlimited model size through memory offloading, Alpa~\cite{alpa} as the model parallelism solver and NVIDIA A100 GPUs as the underlying hardware. Results for DTFM on OPT-65B and Llama-70B are omitted, as the solver exhausts memory due to the prohibitively large state space during search. All baseline methods are evaluated exclusively on edge devices.


\mypar{Cloud-comparable single-GPU performance}
As illustrated in~\autoref{fig:eval-e2e-lat}, under the matched bandwidth and achieved-FLOPS envelopes described above, \sys achieves per-batch runtime comparable to the cloud, while the baseline methods experience slowdowns of up to 15× compared to the cloud setup.
The success of \sys stems from the parameter-server architecture, which reduces uplink communication by at least 3× while keeping downlink communication from becoming the bottleneck.
We do observe a performance gap between \sys and the cloud for small models (\eg OPT-1.3B). In those cases, a cloud deployment can train on a single GPU with enough local memory bandwidth to exceed the aggregate network bandwidth available at the edge. Consequently, \sys incurs 52 seconds of additional runtime (1.5× slower) in these scenarios. Even so, \sys remains up to 10× faster than the baselines.
DTFM cannot further reduce runtime because its communication overhead is effectively fixed: each device must send data equivalent to a layer's size once, leading to runtimes 8--10× longer than cloud training. Alpa's use of TP increases communication volume, resulting in even higher runtimes. In addition, Alpa assigns equal communication and computation workloads to both stragglers and non-stragglers, amplifying the slowdown caused by stragglers.

\begin{table}[t]
	\centering
	\caption{Absolute wall-clock per-batch time (seconds) for representative configurations, derived from the cost model using median edge device parameters (6\,TFLOPS, 55\,MB/s DL, 7.5\,MB/s UL) from \S\ref{sec:background-device} and batch size 128, sequence length 1024. Cloud baseline uses an A100 with memory offloading via PCIe\,4.0.}
	\label{tab:absolute-times}
	\resizebox{\columnwidth}{!}{
		\begin{tabular}{lccc}
			\toprule
			\textbf{Configuration}    & \textbf{Cloud (A100)} & \textbf{\sys} & \textbf{DTFM} \\
			\midrule
			256 devices + OPT-13B     & 33.6\,s               & 37.3\,s       & 3466.7\,s     \\
			512 devices + Llama2-13B  & 33.6\,s               & 16.6\,s       & 3466.7\,s     \\
			1024 devices + Llama2-70B & 180.8\,s              & 30.4\,s       & --            \\
			\bottomrule
		\end{tabular}
	}
\end{table}

Table~\ref{tab:absolute-times} reports the absolute per-batch wall-clock time implied by the same cost model used for normalized comparisons. For example, under 512-device Llama2-13B ($m{=}128\times1024$, $n{=}q{=}5120$, $\alpha{=}\beta{=}10$), one representative attention GEMM level gives $C_{\mathrm{DL}}\approx((\alpha nb)+(n\beta b))/W_{\mathrm{DL}}+L_{\mathrm{DL}}\approx0.0545$\,s, $C_{\mathrm{UL}}\approx0.0107$\,s, and $C_{\mathrm{Comp}}\approx4.4\,\mu$s, so level time is DL-dominated; summing $4\times(h\!\times\!h)$, one $(h\!\times\!H)$, and one $(H\!\times\!h)$ GEMM per layer over $L{=}40$ yields $\approx16.6$\,s. For the cloud baseline, we estimate $T_{\mathrm{A100}}\approx 6N(B\!\cdot\!T)/312\,\mathrm{TFLOPS} + 2N/32\,\mathrm{GB/s}$ (compute + PCIe offload), giving $\approx33.6$\,s for 13B and $\approx180.8$\,s for 70B.

\begin{figure}[t]
	\centering
	\centering
	\includegraphics[width=\linewidth]{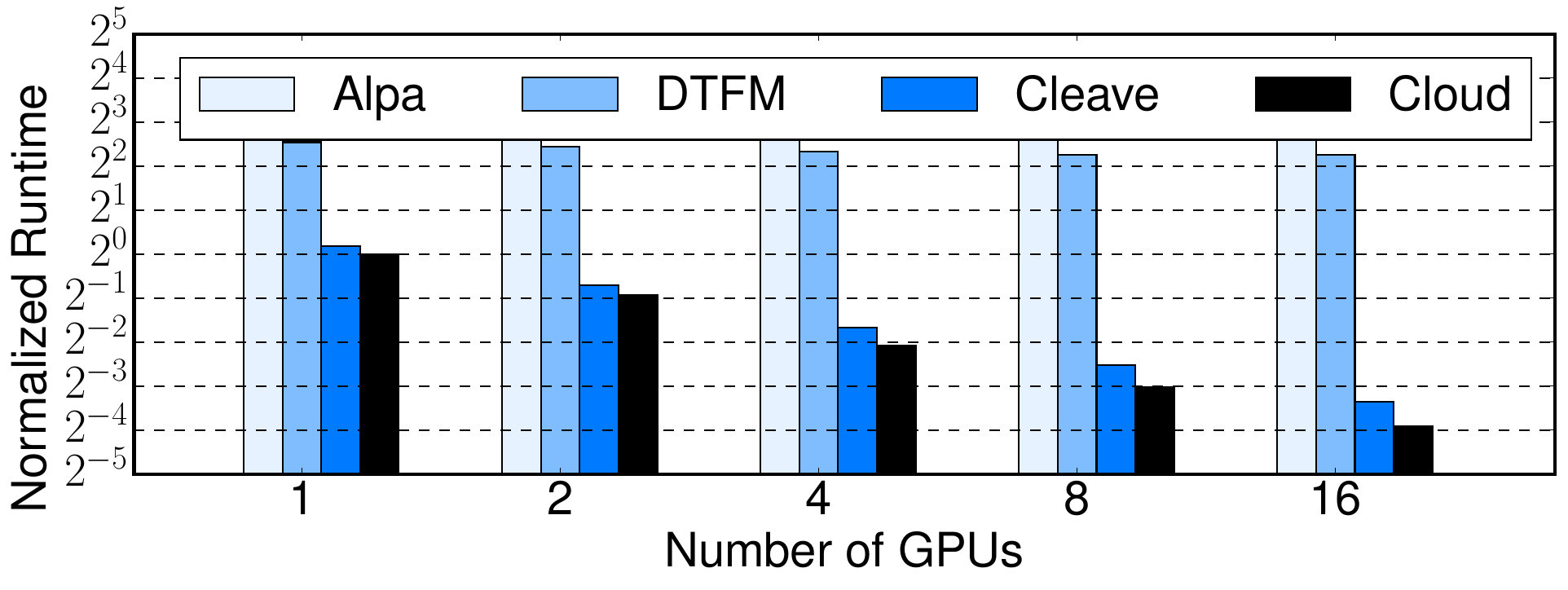}
	\caption{Normalized per-batch runtime for OPT-13B against multiple GPUs. Under the matched-resource methodology of \S\ref{sec:evaluation}, \sys scales with additional edge devices to remain within $2\times$ of multi-GPU cloud performance.}
	\label{fig:eval-e2e-lat-multi}
\end{figure}

\mypar{Cloud-comparable multi-GPU performance}
We next compare \sys against multi-GPU cloud training under the same matched-resource methodology, as shown in~\autoref{fig:eval-e2e-lat-multi}. Based on the device counts used in~\autoref{fig:eval-e2e-lat}, we scale out the number of edge devices proportionally to the number of cloud GPUs used in each system.
Baseline methods, which are primarily limited by AllReduce and AlltoAll communication overhead, fail to benefit from the increased number of devices and therefore exhibit significantly lower performance in this setting. While load balancing becomes slightly more challenging for \sys at larger scales because heterogeneity grows, it maintains per-batch runtime within a 2× margin of multi-GPU cloud setups.

\begin{figure}[t]
	\centering
	\includegraphics[width=.9\linewidth]{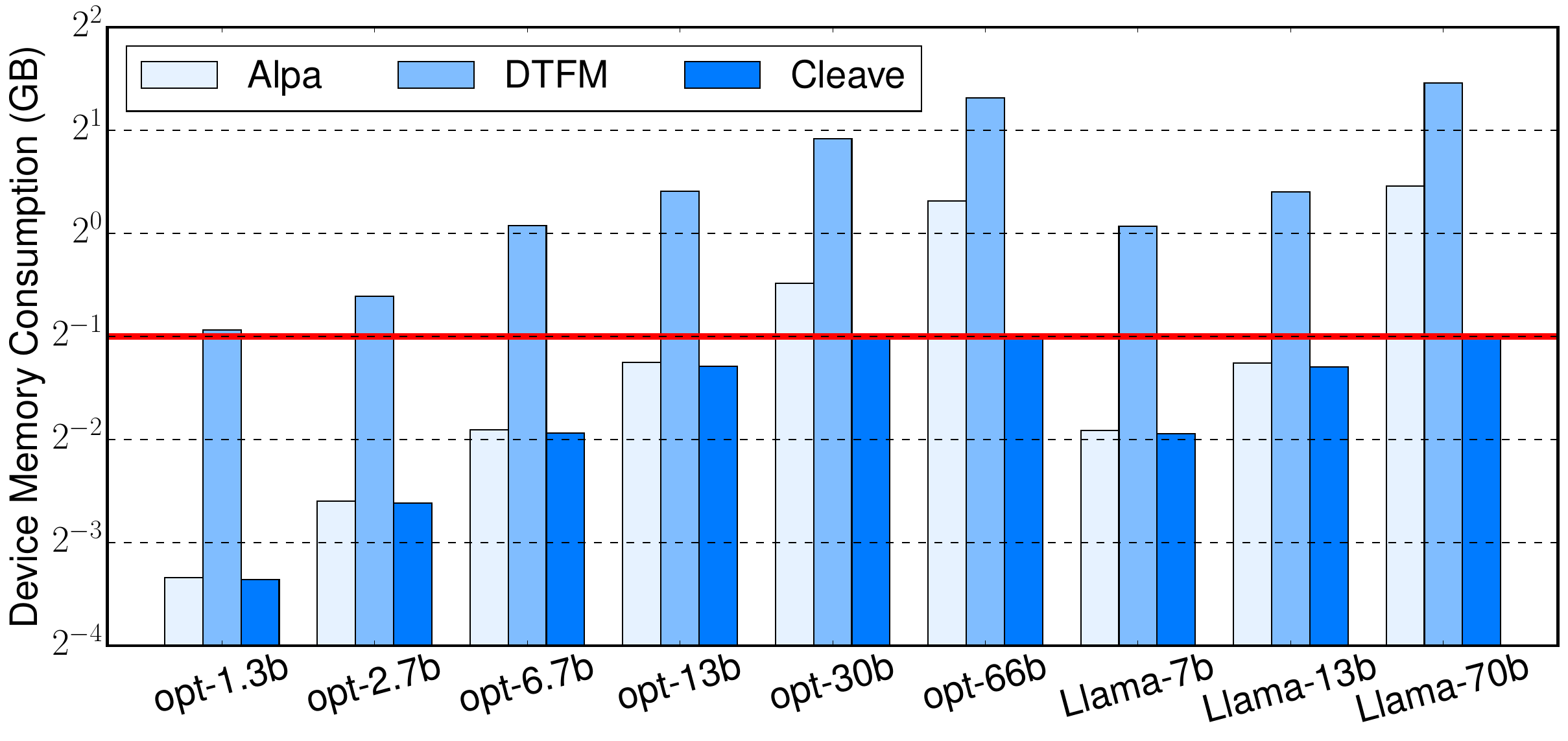}
	\caption{Per-device memory consumption with 8192 candidate devices, where each system can choose how many devices to use. \sys can train large models, while the baselines can encounter out-of-memory failures. The red line marks the 0.5GB mobile-phone capacity~\cite{flexnn}.}
	\label{fig:eval-e2e-memory}
\end{figure}

\mypar{Memory consumption per device}
We evaluate peak memory consumption during training using \sys.
The result is shown in~\autoref{fig:eval-e2e-memory}.
We observe that \sys can scale to all types of models (including those exceeding 30B parameters), with its ability to cap memory consumption at device limit.
The reason is that GEMMs in \sys are partitioned at fine granularity, so the amount of data received can be tuned to each device.
The memory of baselines increases linearly with model size, with DTFM consuming more memory than others.
Although 8192 devices are provided, DP+PP in DTFM allows a maximum of 4096 devices to be used for OPT-1.3B, still resulting in larger memory consumption for each layer.
Alpa can reduce this with TP, but for models greater than 30B parameters, the large optimizer size and intermediate size,  which has to stay on device, still exceeds the device capability.
Alpa, in such a case, needs two times more devices to support the same size model as \sys.

\subsection{Handling Stragglers and Device Churn}

\begin{figure}[t]
	\centering
	\includegraphics[width=.85\linewidth]{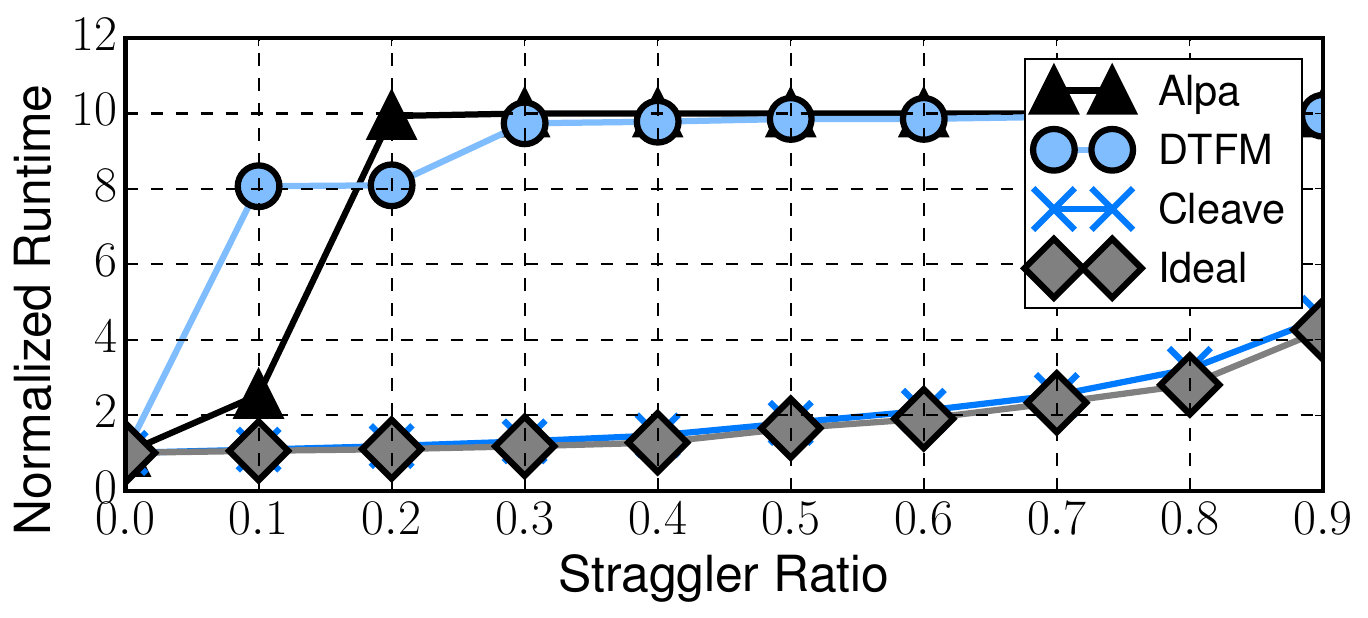}
	\vspace{-0.1in}
	\caption{Per-batch runtime under increasing straggler fractions, normalized to the no-straggler case for each system.}
	\label{fig:eval-straggler}
\end{figure}

\mypar{Impact of stragglers}
We vary the proportion of stragglers among all devices and evaluate training with 32 devices on the OPT-13B model.
The stragglers are set to be 10X slower than the average performance of other devices, including both computation and communication.

The results in~\autoref{fig:eval-straggler} show that \sys is substantially less affected by stragglers than the baselines.
\sys redistributes tasks to non-straggler devices using its cost model. Although runtime still increases as more stragglers appear because fewer devices contribute useful work, \sys remains efficient. On average, it deviates by only 5\% from the ideal case in which straggler workload can be redistributed at the finest possible granularity. Exact load-balanced redistribution is not always achievable because the basic unit of work is a row--column pair.

For baseline methods, the presence of stragglers leads to a 10× slowdown when 20\% of the devices are stragglers. Alpa assigns tasks evenly across all devices, while DTFM involves stragglers in the AllReduce operation in DP. Since training is synchronized, every operation must wait for the slowest device to complete, significantly impacting runtime.

We use OPT-13B as the representative straggler case. Under the same normalization to each system's no-straggler runtime, the qualitative trend should carry to larger OPT and Llama2 models because both communication and computation demand scale proportionally with model size.

\begin{figure}[t]
	\centering
	\includegraphics[width=.9\linewidth]{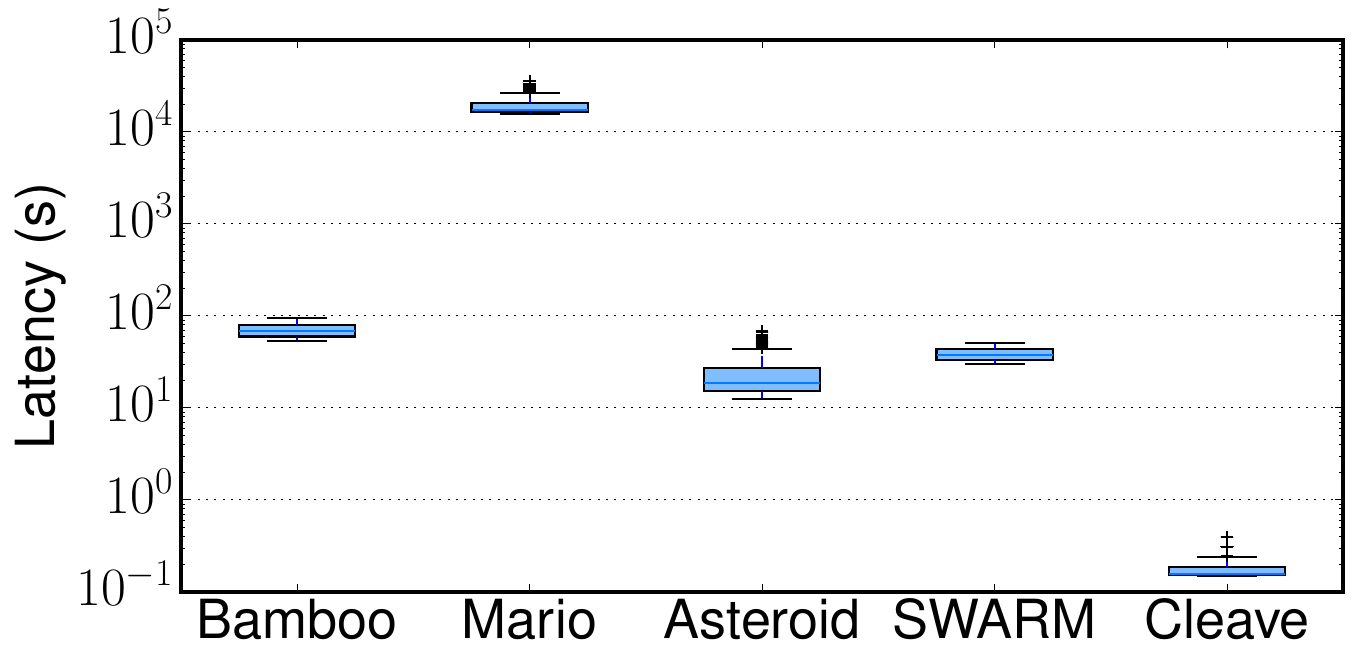}
	\vspace{-0.1in}
	\caption{Absolute latency for recovery from device failure/departure, using OPT-13B with batch size 128 and sequence length 1024.}
	\label{fig:eval-lat-fail}
\end{figure}

\begin{table}[t]
	\centering
	\begin{tabular}{|c|c|c|c|}
		\hline
		Designs           & Comm  & Memory & Runtime \\
		\hline
		\sys              & 0.4GB & 267MB  & 1037s   \\
		\hline
		w/o TP            & 273\% & 576\%  & 413\%   \\
		w/o PS            & 342\% & 121\%  & 543\%   \\
		w/o heterogeneity & 121\% & 100\%  & 325\%   \\
		\hline
	\end{tabular}
	\captionof{table}{Ablation-study results showing the contribution of \sys's components relative to the complete system. We report communication volume (Comm), device memory consumption (Memory), and per-batch runtime for Llama2-13B with batch size 128, sequence length 1024, and 1024 devices.}
	\label{tab:system-breakdown}
\end{table}

\mypar{Impact of device churn}
We investigate the time taken to recover from a device failure. We use OPT-13B with 256 devices for this experiment; other models and device counts exhibit similar trends.
\autoref{fig:eval-lat-fail} shows the recovery time.
Checkpoint-restore-based methods (\eg Mario) are the slowest because they require downloading tens of gigabytes of activation data over constrained links, which takes longer than a single training step.
Other baselines (\eg Bamboo, SWARM, Asteroid) require full recomputation of at least one model layer, along with transmission of the corresponding hidden states. On edge devices with limited compute capacity, such recomputation typically takes around 50 seconds.
In contrast, \sys only needs to transmit and recompute a shard of a GEMM operation, which is approximately 20× smaller than a full model layer. Moreover, this recomputation is distributed across all devices rather than assigned to a single device, leading to recovery that is at least 100× faster.
\review{This speedup is a direct consequence of \sys's fine-grained sharding design, which deliberately minimizes the blast radius of each device failure. Under realistic edge churn of 1\% per hour (10 failures/hr across 1{,}000 devices), each 60\,s training batch encounters ${\sim}0.17$ failures on average. \sys's short incremental recovery path incurs ${<}0.3\%$ overhead per batch, sustaining 99.7\% effective throughput. In contrast, layer-recomputation baselines (Bamboo, SWARM, Asteroid) lose ${\sim}14\%$ throughput per batch, while checkpoint-restore methods (Mario) stall training entirely when recovery exceeds the batch interval.} This analysis assumes statistically independent device failures; correlated events such as network partitions or time-of-day mass disconnects would increase per-event recovery scope and remain important future work.
We also evaluate the time required to enable a new device to join the training process.
\sys enables seamless integration of new devices without pausing training, unlike baselines that incur latency due to layer resharding and weight transfer.

\subsection{Ablation Study}

We study the contributions of TP, the PS architecture, and the cost model independently.
\autoref{tab:system-breakdown} presents the ablation study averaged across all models.



\mypar{Tensor parallelism (TP)}
Eliminating TP increases per-device communication volume by 273\% and runtime by 413\%. Without TP, each device must receive a full matrix rather than rows and columns, and vector-matrix multiplication (GEMV) exposes much less useful asymmetry than GEMM.

\mypar{Parameter server architecture (PS)}
Replacing the PS architecture with peer-to-peer communication results in a 342\% increase in communication volume and a 543\% increase in runtime. Peer-to-peer approaches require broadcasting model parameters, matrix reshaping, and AllReduce operations across devices, all of which substantially increase communication overhead. The PS architecture avoids this by centralizing parameter storage and using optimized downlink (DL) transmission, which utilizes bandwidth more effectively and lowers runtime.
The memory increase arises because the optimizer must now be loaded on devices rather than kept on the PS.

\mypar{Optimizer accounting}
For GEMM-offloaded training in \sys, PS-side optimizer updates execute on the host CPU and are largely hidden by DAG-level pipelining behind backward GEMM computations. As detailed in \S\ref{sec:practical-concerns}, the exposed tail remains negligible relative to 60--120\,s batch times and does not create a net penalty versus host-offloaded cloud baselines such as DeepSpeed ZeRO-Offload~\cite{zero-offload}.

\mypar{Device heterogeneity awareness}
Removing device-heterogeneity awareness from GEMM scheduling increases runtime by up to 325\% and communication volume by 21\%. Without load balancing, stragglers receive the same workload as faster devices and therefore delay synchronization. Moreover, model parameters are replicated to more devices instead of being skewed toward stronger devices, which further increases communication volume.



\subsection{Training Scalability}
\begin{figure}[t]
	\centering
	\includegraphics[width=\linewidth]{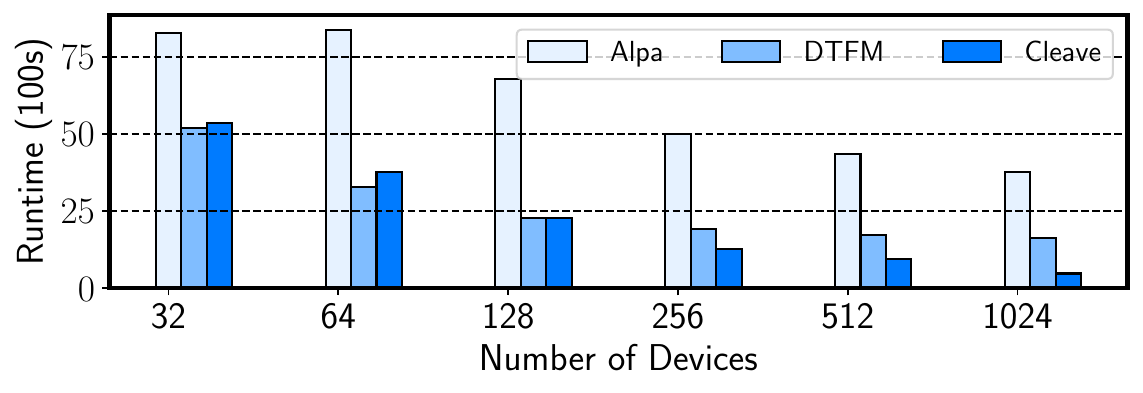}
	\vspace{-0.25in}
	\caption{Per-batch runtime of OPT-13B when scaling the number of devices at a fixed batch size (steeper decline is better).}
	\label{fig:eval-num-device-opt-13b}
\end{figure}

We explore both weak scaling, \ie scaling model size or batch size together with the number of devices, and strong scaling, \ie fixing model size while scaling the number of devices.


\mypar{Number of Devices}
We evaluate scaling with varying device counts under a fixed model and input size, \ie strong scaling.
As~\autoref{fig:eval-num-device-opt-13b} shows, \sys reduces per-batch runtime effectively while scaling to up to 8× more devices. DTFM, however, struggles to scale because its communication cost remains effectively constant: when the number of devices doubles, the per-device communication volume does not fall because of model-parameter AllReduce. As the number of devices increases from 32 to 64 in~\autoref{fig:eval-num-device-opt-13b}, DTFM's training time even increases. In this regime, where there are more devices than model layers, DTFM relies on DP+PP, which raises communication by transmitting additional model parameters.
Alpa also scales poorly and has higher runtime than \sys. Regardless of device speed, Alpa assigns the same load to all devices, so runtime is bounded by the slowest participant. When the number of devices doubles, \sys reduces runtime by 1.8×, whereas Alpa achieves only a 1.3× reduction. \sys's near-linear scaling comes from the load-balancing strategy in its cost model, which assigns less work to weaker devices. The remaining overhead in \sys comes from sequential dependencies between GEMMs.

\begin{figure}[t]
	\centering
	\includegraphics[width=\linewidth]{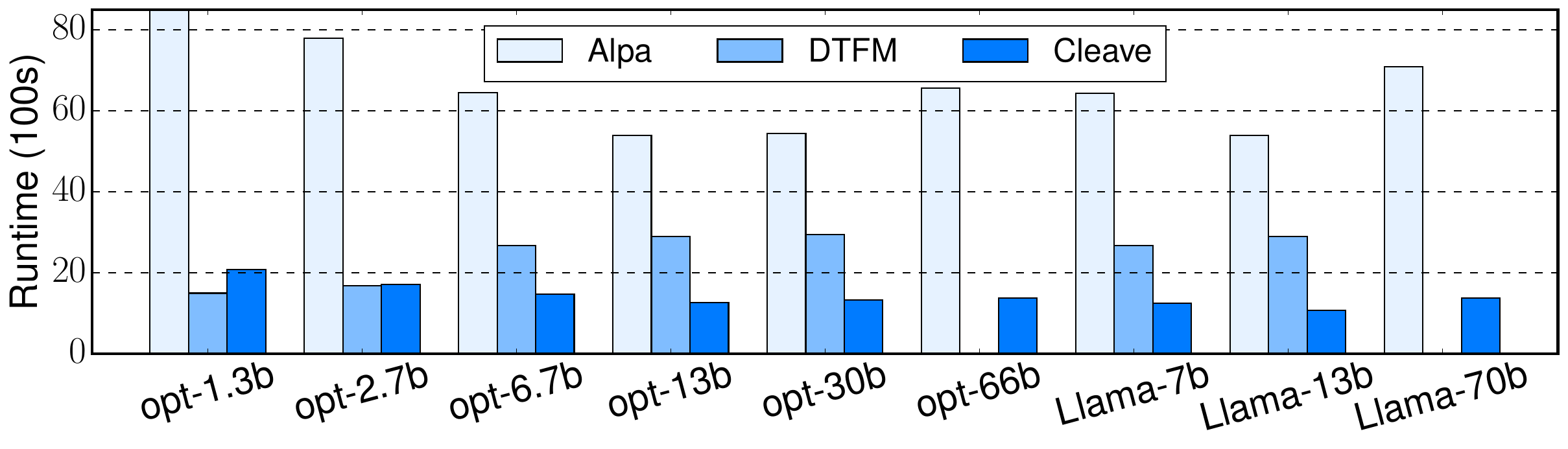}
	\vspace{-0.25in}
	\caption{Per-batch runtime when scaling model size proportionally with the number of devices (flatter is better).}
	\label{fig:eval-model-size}
\end{figure}

\mypar{Model size}
We fix the sequence length (1024) and batch size (128) while varying model size. The number of devices scales proportionally with model size, with the 70B model mapped to 1024 devices. As shown in~\autoref{fig:eval-model-size}, \sys maintains nearly constant runtime across all settings, demonstrating effective weak scaling. Because the same amount of work per device produces a similar communication pattern, runtime remains stable. In contrast, DTFM fails to scale to 1024 devices for OPT-66B and Llama-70B because its memory requirements exceed 1TB, surpassing server capacity. Alpa's uniform GEMM assignment creates stragglers and hinders scaling.

\begin{figure}[t]
	\centering
	\includegraphics[width=\linewidth]{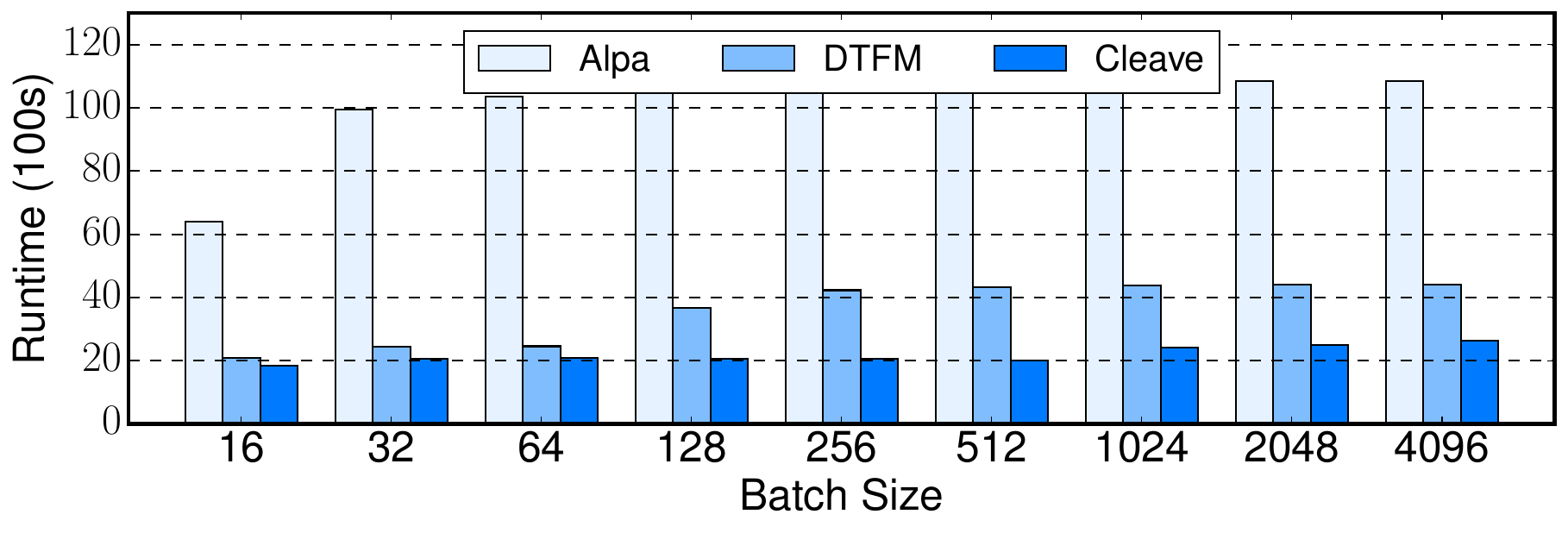}
	\vspace{-0.25in}
	\caption{Per-batch runtime of OPT-13B when scaling batch size proportionally with the number of devices (flatter is better).}
	\label{fig:eval-batch-size}
\end{figure}

\mypar{Batch size}
We fix the model size (OPT-13B) and sequence length while varying batch size, with each device processing a mini-batch size of 2. \autoref{fig:eval-batch-size} shows that \sys again maintains nearly constant runtime, indicating that the cost model uses both computation and bandwidth efficiently. DTFM scales better at smaller batch sizes (16--64) because PP keeps communication volume low. However, once batch size reaches 128 and beyond, DP becomes necessary and increases communication volume on each device. Alpa still has about 7× longer runtime for the same reasons that it fails to scale with model size.

\section{Discussion on Practical Concerns}
\label{sec:practical-concerns}

We first address four deployment questions for \sys: coordinator-side cost, the operating range of a single parameter server, how the system scales beyond one parameter server, and whether PS-side optimizer execution changes the comparison to cloud training. For each concern, we tie the deployment question to the aspect of \sys's design that matters in practice. We next discuss likely deployment settings and broader considerations including energy efficiency, robustness, and support for active devices.

\mypar{Cost considerations}
A first question is whether \sys lowers deployment cost or merely moves the bill from GPUs to another part of the system. In the setting we target, participant edge devices contribute spare capacity, so the cloud-side role shrinks from a multi-GPU trainer to a CPU-only coordinator. Table~\ref{tab:cost-comparison} compares coordinator-side cost under equal runtime. Using AWS on-demand pricing~\cite{aws-ec2-pricing}, it shows a reduction of about {4.9$\times$} relative to on-demand 8$\times$A100 cloud training and {6.2$\times$} relative to the larger A100 configuration. We intentionally scope this comparison to institution-hosted or self-hosted deployments by excluding network-egress charges~\cite{edge-cost-survey}, so the takeaway is coordinator-side savings when opt-in spare edge resources are available, not a claim of universal cost superiority.

\begin{table}[t]
	\centering
	\caption{Equal-runtime infrastructure cost comparison between cloud GPU training and \sys.
		All hourly prices are on-demand. The \sys column counts only the coordinator-side cloud cost; edge devices are treated as opt-in spare resources.}
	\label{tab:cost-comparison}
	\resizebox{\columnwidth}{!}{
		\setlength{\tabcolsep}{3pt}
		\begin{tabular}{lcccccc}
			\toprule
			\textbf{System} & \textbf{Instance} & \textbf{Accelerator} & \textbf{GPU mem.} & \textbf{Host mem.} & \textbf{\$/hr} \\
			\midrule
			Cloud           & p4d.24xlarge      & 8$\times$A100        & 320\,GB           & 1152\,GiB          & \$21.96        \\
			Cloud           & p4de.24xlarge     & 8$\times$A100        & 640\,GB           & 1152\,GiB          & \$27.45        \\
			Cloud           & p5.48xlarge       & 8$\times$H100        & 640\,GB           & 2048\,GiB          & \$55.04        \\
			\midrule
			\sys            & m6in.16xlarge     & 64 vCPU              & --                & 256\,GiB           & \$4.46         \\
			\bottomrule
		\end{tabular}
	}
\end{table}

\mypar{Single-PS operating envelope}
A first scalability question is how far a single parameter server can go before coordination dominates. In \sys, service demand is driven by \emph{per-level GEMM payloads}, not by naively summing every device's peak link rate, because the PS serves one DAG level at a time and overlaps that service with seconds-scale device-side GEMM execution. With a realistic 200~Gbps CPU-only PS and typical fixed-broadband downlink speeds of 250~Mbps~\cite{speedtest}, a single server can therefore support roughly $10^3$ devices, i.e., about 1,000--2,000 concurrent participants. For a representative QKV projection GEMM ($4096\times4096$) with 1,000 devices each receiving approximately four rows and four columns (FP16), per-device downlink is only about 65\,KB and aggregate per-GEMM downlink is about 65\,MB, which is served in about 2.6\,ms at 25\,GB/s; likewise, 1,000 devices at 5--10\,MB/s uplink remain within a 25\,GB/s PS budget. For the deployment range we target, this keeps a single PS within budget while the cost model's per-device constraints ($W_k^d$ and $W_k^u$ in \autoref{sec:gemm_scheduling}) remain the first-order design rule.

\mypar{Multi-PS scale-out}
A separate question is what happens once sustained demand moves beyond that operating envelope. Here the important observation is that \sys already decomposes work by DAG level and device subset, so scaling out does not require a new training abstraction. In the single-PS case, heterogeneous completion times can create bursty aggregate demand, which the PS smooths through staggered dispatch and flow-control backpressure. Beyond that point, the same design extends naturally to multi-PS sharding and replication; with $N$ balanced PS instances, per-PS demand decreases approximately as $1/N$, consistent with distributed PS techniques such as Beldi~\cite{beldi}. We therefore scale beyond one PS once sustained per-level demand approaches service capacity, rather than treating multi-PS execution as a different system regime.

\mypar{Optimizer step accounting and pipelining}
Another fairness question is whether placing Adam on the PS CPU adds overhead that the cloud baseline does not pay. Two observations keep that from changing the comparison. The first is architectural: our single-GPU cloud baseline also executes optimizer updates in host memory, so PS-side optimizer traffic is not a differential penalty. The second is scheduling: \sys pipelines optimizer work by DAG level during backward execution instead of exposing it as a monolithic post-pass. For Llama2-13B, the total optimizer traffic is approximately $13\text{B}\times(2{+}2{+}2{+}8{+}8{+}4)$ bytes/parameter $\approx 338\,\text{GB}$, which would take about $338/150=2.25\,\text{s}$ at host DDR5 bandwidth of $150\,\text{GB/s}$ if exposed in full. In practice, 40 transformer layers reduce this to $338\,\text{GB}/40=8.45\,\text{GB}$, i.e., about $56\,\text{ms}$ per layer; because per-layer backward GEMM time remains in seconds, updates for layers $L\rightarrow2$ are hidden and only the final layer can be exposed. The visible overhead is therefore small and does not alter the overall cloud comparison.

\mypar{Prospective users of \sys}
\sys is not intended as a drop-in replacement for every cloud-training workflow. It is most compelling when an organization already controls decentralized spare compute that is fragmented, intermittent, and otherwise unusable for synchronous training. That points to two natural user groups: (i) organizations that can pool opt-in idle phones, laptops, or other edge devices instead of renting large cloud GPU clusters, and (ii) operators seeking to harvest time-varying spare compute from AI-RAN deployments. In both cases, \sys turns that fragmented capacity into usable synchronous training throughput through fine-grained sharding, heterogeneity-aware scheduling, and the ability to contract or evict work when radio demand rises.

\mypar{Training data distribution}
Data placement is another practical concern. Because \sys already uses the PS to dispatch forward inputs, data can be pre-distributed to devices or streamed alongside model weights. In our evaluated setting, the PS holds the dataset and streams batch embeddings as part of the forward-pass downlink dispatch, and that traffic is already captured in the downlink cost term of the model (\S\ref{sec:gemm_scheduling}). For public datasets, PS-side streaming is therefore practical; for privacy-sensitive deployments, data can be pre-distributed before training begins. Non-IID data placement, a common federated-learning concern, is outside our scope because \sys targets synchronous full-batch training rather than per-device local gradient accumulation.

\mypar{Energy consumption and carbon footprint}
Energy efficiency matters only if edge-assisted training reduces total system impact rather than merely shifting power draw off the cloud. Under the assumptions in our companion analysis~\cite{edgetrain}---opt-in spare devices, fixed-site charging, and amortized embodied carbon---replacing cloud GPU compute with already-provisioned edge compute changes both the operational and embodied cost profile. Under those assumptions, decentralized edge training remains 1.5--5$\times$ more energy efficient than cloud GPU training, with carbon-footprint reductions of 6$\times$ for smartphone-class devices and 3.5$\times$ for laptop-class devices under representative settings (e.g., 10~MB/s per device and 0.5~W peak WiFi power). This aligns with \sys's design: because total computation and communication volume remain fixed as the number of participating devices grows, shifting execution from data-center GPUs to amortized edge devices yields 4--8$\times$ lower total carbon footprint in that analysis.

\mypar{Architecture centralization}
The parameter server introduces a familiar trade-off between simpler coordination and centralized control. In \sys, that centralization is also what makes the system practical: it coordinates heterogeneous devices, avoids peer-to-peer collectives that are poorly matched to edge links, and handles churn through a single scheduler. We therefore use a single PS as the pragmatic default for moderate-scale deployments. When the coordination dependency becomes the dominant scale or availability risk, the multi-PS extension in \S\ref{sec:gemm_scheduling} provides the path to distribute control.

\mypar{Parameter server fault tolerance}
That same centralization means a PS failure halts training. We view this as an operational concern at the coordination layer, not as a limitation of \sys's sub-GEMM sharding model itself. The natural mitigation is standard checkpoint/restart of model parameters and optimizer state to persistent storage (e.g., every $N$ batches), coupled with automatic recovery on a standby instance. In larger deployments, the multi-PS extension (\S\ref{sec:gemm_scheduling}) adds redundancy: with $N$ balanced PS instances, a single failure affects only $1/N$ of the device fleet while the remaining instances continue training their assigned layers.

\mypar{Privacy considerations}
Because the PS receives intermediate activations from all devices, privacy-sensitive deployments must consider gradient-inversion risk. For our primary target setting---training on public datasets---this concern is substantially weaker than in privacy-sensitive federated deployments. When stronger privacy guarantees are needed, \sys can be composed with standard mitigations such as differential privacy or secure aggregation, at the cost of additional communication overhead. Those protections operate over the same PS-mediated data path and are therefore orthogonal to \sys's core sharding contribution.

\mypar{Robustness to poisoning attacks}
Malicious workers could also return incorrect partial GEMM results and silently corrupt training. Here the PS's position in the loop is useful: because it dispatches the inputs and receives the returned block, it can verify algebraic consistency before accepting a contribution. Concretely, for matrix multiplication $C = AB$, \sys can sample random vectors $r, s \in \mathbb{R}^n$ and verify whether $r^\top (AB) s = (Ar)^\top (Bs)$~\cite{DBLP:journals/csur/MotwaniR96}. This probabilistic check detects even single-entry corruption with high probability while incurring only $\mathcal{O}(n)$ overhead, where $n$ is the largest dimension of $A$ or $B$; because vectors are freshly generated at runtime and the check reduces to lightweight GEMV operations, the false-negative probability is at most $\mathcal{O}(2^{-n})$ and the mechanism remains practical even against white-box adversaries on modern CPUs~\cite{moe-gen}.

\mypar{Adaptation to active devices}
Our current target is idle, charging, network-connected devices, but a natural question is how \sys would behave once devices become active again. Fine-grained sub-GEMM assignments help here because disruption stays local: foreground activity, thermal events, or short network jitter delay only a few shards rather than an entire layer or batch. This yields a two-level adaptation strategy. At the \emph{micro level}, transient interference delays only the affected shards while other devices continue to make progress; at the \emph{macro level}, longer-lived changes in user activity can be handled by periodically refreshing each device's effective capability from runtime telemetry and redistributing work away from chronically degraded devices while re-admitting recovered ones.

\mypar{Sequence length sensitivity}
All experiments in this paper use sequence length 1024, so an important open question is how the trade-off shifts at the 4K--128K contexts used in modern LLM training. Longer sequences do not merely scale runtime uniformly; they specifically amplify attention GEMM communication as a function of $s$, which changes the compute-to-communication balance and can shift the regimes in which \sys is most attractive. The crossover conditions derived in the Appendix already expose this dependence on $s$, and they suggest that \sys's uplink advantage may grow for longer sequences because the output shard size ($\alpha\beta b$) is sequence-independent while the attention GEMM input ($Bsh$ activations) grows linearly. Evaluating \sys at contemporary long-context settings is therefore important future work.

\section{Conclusions}
We have presented \sys, a framework for training foundation models by harnessing idle edge compute while accounting for the core characteristics of the edge environment, including memory constraints, compute/network heterogeneity, and dynamic availability.
\review{At its core, \sys exploits a structural alignment between the asymmetric I/O pattern of GEMM operations---where inputs dispatched over downlink are substantially larger than outputs returned over uplink---and the asymmetric link speeds of edge networks, enabling a parameter-server-centric architecture where per-device communication \emph{decreases} with scale rather than remaining constant.}
To this end, \sys introduces a selective hybrid tensor-parallelism technique and builds on a parameter-server-based training framework to support scalable and efficient workload distribution. \sys explicitly addresses stragglers and handles device churn through fast recovery and seamless integration of new devices, ensuring robustness throughout training.
Our findings indicate that \sys is effective for large-scale distributed foundation model training with edge devices and, in our evaluated setting, achieves cloud-comparable per-batch runtime, underscoring its potential for democratizing foundation model development.

\printbibliography

\appendix


\section{Communication Efficiency: Homogeneous}

We analyze the per-device communication volume and derive conditions under which \sys achieves superior communication efficiency compared to conventional parallelism strategies.
We begin with a homogeneous setting where all devices have identical workloads, FLOPS, and bandwidth—a configuration that favors the baseline DP+PP+TP parallelism—and subsequently extend to heterogeneous environments with stochastic latency models.

\subsection{3D Parallelism Communication Analysis}

We first establish the communication volume in conventional DP, PP, TP settings.

\mypar{Volume for data parallelism}
In data parallelism, the AllReduce communication volume is $\frac{B}{b_\mu}(4h^2 + 3hH)L$, as each transformer layer's attention mechanism involves four weight matrices (Q, K, V, O), each of dimension $h \times h$, contributing $4h^2$ parameters. The Llama architecture~\cite{llama2} employs three matrices in the MLP layer (up projection, gate projection, and down projection), contributing $3hH$ parameters.
This formulation is consistent with the Megatron-LM framework analysis~\cite{megatron}, where gradient synchronization scales linearly with parameter count. The number of model replicas in DP is $\frac{B}{b_\mu}$, requiring each DP stage to transmit its gradients.

\mypar{Volume for pipeline parallelism}
Pipeline parallelism introduces additional communication between stages, amounting to $2(p{-}1)Bsh$ for forward and backward propagation, given $p \leq L$ pipeline stages.

\mypar{Volume for tensor parallelism}
loud-style tensor parallelism involves AllReduce operations for intermediate results, adding $4tBshL$ of communication for MLP and attention layers in both propagation directions, where $t$ denotes the tensor parallel degree.

The minimal per-device communication volume under conventional 3D parallelism is therefore:
\begin{equation}
	V_{\text{baseline}} = \frac{(4h^2 + 3hH)}{t} + 2 \cdot \mathds{1}{p>1} \cdot Bsh + 2 \cdot \mathds{1}{t>1} \cdot Bsh \label{eq:volume-baseline}
\end{equation}

where the total number of devices is $D = tp\frac{B}{b_\mu}$. Communication volume is symmetric for both uplink and downlink in conventional approaches, a property that fails to exploit the bandwidth asymmetry characteristic of edge environments.

\begin{table}[t]
	\centering
	\resizebox{\linewidth}{!}{
		\begin{tabular}{r|lr|l}
			\hline
			$a$     & number of attention heads   & $s$                 & sequence length                 \\
			$b_\mu$ & microbatch size             & $t$                 & tensor parallel size            \\
			$h$     & hidden dimension size       & $B$                 & batch size                      \\
			$p$     & pipeline parallel size      & $L$                 & number of transformer layers    \\
			$H$     & intermediate dimension size & $\mathds{1}_{cond}$ & 1 if $cond$ is met, 0 otherwise \\
			\hline
		\end{tabular}
	}
	\caption{Variable names following Megatron convention~\cite{megatron}.}
	\label{tab:megatron-var-names}
\end{table}

\subsection{\sys Communication Volume}

In \sys, we first consider the case without caching optimizations or operator fusion. The communication volume directly related to model weights is $(8Bsh^2 + 18BshH)L$ for QKVO projection and MLP layers respectively, accounting for both forward and backward propagation. Additionally, attention weights and outputs require $4Bs^2hL$ of communication. For each device, the communication volume equals the total divided by $D$.

Setting $H = 4h$ as is standard in transformer architectures following the original design rationale~\cite{attention}, \sys achieves lower downlink communication volume than baselines under the condition:
\begin{equation}
	D > \frac{3(80 + 4s)L}{16h/(tBs) + 4} \tag{7}
\end{equation}

This condition characterizes downlink-bounded communication, which is less common than uplink-bounded scenarios in edge environments where uplink bandwidth is typically 2--10$\times$ lower than downlink~\cite{speedtest}.

For uplink communication, without caching or further optimizations, \sys requires transmitting all model parameters $(4h^2 + 3hH)L$, intermediate results $BshL$, and additional activations $(2BsH + 5Bsh + Bs^2h)L$ for MLP, QKVO, and attention weights respectively. \sys achieves uplink communication benefits when:
\begin{equation}
	D > \frac{(8h/(Bs) + 13 + s)L}{8h/(tBs) + 2} \label{eq:uplink-benifits}
\end{equation}

This demonstrates that \sys's communication advantages are most pronounced in uplink-constrained environments, which is precisely the characteristic of edge network deployments.

\subsection{Tightening Bounds of Communication}

The bounds in~\autoref{eq:volume-baseline} and~\autoref{eq:uplink-benifits} compare aggregate communication volumes, providing sufficient but not necessary conditions. We derive tighter characterizations by analyzing the temporal structure enabled by \sys's streaming protocol.

The streaming mechanism (illustrated in~\autoref{fig:transformer-dag}) enables overlap between downlink transmission, computation, and uplink transmission. Let $T_{\text{DL}}$, $T_{\text{comp}}$, and $T_{\text{UL}}$ denote the time for downloading, computing, and uploading one row-column pair respectively. For a device processing $k$ pairs, the effective completion time follows the pipeline model~\cite{DBLP:books/daglib/0028244}:
\begin{equation}
	T_{\text{pipeline}}(k) = T_{\text{DL}} + (k{-}1) \cdot \max(T_{\text{DL}}, T_{\text{comp}}, T_{\text{UL}}) + T_{\text{comp}} + T_{\text{UL}} \tag{9}
\end{equation}

This captures the pipeline fill phase ($T_{\text{DL}}$), steady-state execution at the rate of the slowest stage, and drain phase ($T_{\text{comp}} + T_{\text{UL}}$). The refined crossover condition compares this pipelined makespan against the baseline's AllReduce latency, which scales as $O(\alpha \cdot \lceil \log_2 D \rceil)$ for ring-based implementations~\cite{DBLP:journals/ijhpca/ThakurRG05}, where $\alpha$ is the per-message latency.

The DAG structure (\autoref{fig:transformer-dag}) imposes $S$ synchronization barriers corresponding to levels with memory dependencies. The total \sys execution time decomposes as:
\begin{equation}
	T = \sum_{s=0}^{S-1} \left( T_{\text{GEMM}}(s) + T_{\text{sync}}(s) \right) \tag{10}
\end{equation}

where $T_{\text{sync}}(s)$ captures waiting time at level $s$. In homogeneous settings, $T_{\text{sync}} \approx 0$ under optimal load balancing. Including this term explicitly enables principled extension to heterogeneous analysis in later section.

Combining Equations (9) and (10), the tightened condition for \sys advantage becomes:
\begin{equation}
	D > \frac{S \cdot T_{\text{pipeline}}(W/D)}{\alpha \cdot \lceil \log_2 D \rceil + \beta \cdot V_{\text{baseline}}/W_d} \tag{11}
\end{equation}

where $W$ is the total workload per level, $\alpha$ and $\beta$ are latency and bandwidth coefficients respectively, and $W_d$ is the downlink bandwidth. This bound is tighter than Equations (7)--(8) by a factor of $O(\log D)$ in typical configurations.

\section{Communication Efficiency: Heterogeneous}

The heterogeneous setting requires fundamentally different analytical tools because the optimization problem—minimizing makespan subject to memory and workload constraints—is a variant of the unrelated parallel machine scheduling problem, known to be NP-hard~\cite{lenstra1990approximation}.

\subsection{Problem Formulation and Complexity}

The workload allocation problem in \sys can be formulated as a generalized assignment problem. Let $x_{ik}$ denote the fraction of GEMM $i$ assigned to device $k$, with constraints $\sum_k x_{ik} = 1$ for all $i$ and $x_{ik} \geq 0$. The objective is to minimize the maximum completion time (makespan) across devices:

\begin{equation}
	\min_{x} \max_{k} \sum_i \frac{x_{ik} \cdot W_i}{F_k} + C_{\text{COMM}}(x, k)
	\tag{17}
\end{equation}

subject to memory constraints $\sum_i x_{ik} \cdot M_i \leq M_k$ for all devices $k$.

This problem admits a $(2 - \frac{1}{m})$-approximation via the Longest Processing Time (LPT) heuristic~\cite{graham1969bounds}, where $m$ is the number of machines. However, \sys's structure enables better guarantees.

\subsection{Exploiting Structural Properties}

\sys's DAG structure provides two properties that enable tighter analysis: GEMMs within a level are independent (no memory dependencies), and workload is arbitrarily divisible at the row-column granularity. Under these conditions, the optimal makespan at each level satisfies the lower bound:

\begin{equation}
	T^*_{\text{level}}(s) \geq \max\left( \frac{\sum_i W_i(s)}{\sum_k F_k}, \max_i \frac{W_i(s)}{F_{\max}} \right)
	\tag{18}
\end{equation}

where $W_i(s)$ is the workload of GEMM $i$ at level $s$, $F_k$ is device $k$'s compute capability, and $F_{\max} = \max_k F_k$. The first term represents the parallelism-limited bound (total work divided by total capacity), while the second represents the serialization-limited bound (largest indivisible unit).

The Gurobi solver~\cite{gurobi} achieves a makespan within a factor of $(1 + \epsilon)$ of this lower bound for any $\epsilon > 0$ given sufficient solver time, leveraging the convexity of the continuous relaxation and the effectiveness of branch-and-bound for the integrality constraints.

\subsection{Stochastic Performance Bounds}

When device capabilities are drawn from a distribution, as in our evaluation using AI-Benchmark data~\cite{ai-benchmark}, we derive expected-case bounds using order statistics theory~\cite{david2004order}.

Let $F_k \sim \mathcal{F}$ with mean $\mu_F$ and variance $\sigma^2_F$. The coefficient of variation $c_v = \sigma_F / \mu_F$ characterizes heterogeneity. For a load-balanced allocation where each device receives workload proportional to its capability, the expected makespan scales as:

\begin{equation}
	\mathbb{E}[T_{\text{hetero}}] \approx T_{\text{homo}} \cdot \left(1 + \frac{c_v^2}{2} \cdot g(D)\right)
	\tag{19}
\end{equation}

where $g(D)$ is a decreasing function capturing the load-balancing benefit of additional devices.

For \sys's fine-grained allocation at row-column granularity, concentration inequalities~\cite{boucheron2013concentration} yield $g(D) \approx 1/\sqrt{D}$. This reflects the law of large numbers: with many small tasks, deviations from optimal balance average out. In contrast, for coarse-grained baselines like DTFM where the minimum allocation unit is a full layer, $g(D) \approx 1$, providing no asymptotic improvement from additional devices.

This analysis explains the empirical observation (\autoref{fig:eval-straggler}) that \sys deviates by only 5\% from the ideal case under straggler presence, while baselines experience $10\times$ slowdowns.

\section{Distributional Latency Modeling}

Network latency in edge environments exhibits heavy-tailed behavior due to variable wireless conditions, congestion, and device-level scheduling artifacts. The cost model in Section 3.2 treats latency as deterministic constants $L^d_k$ and $L^u_k$, which underestimates tail effects that dominate synchronous training performance.

\subsection{Distributional Model}

Empirical studies of mobile network latency~\cite{DBLP:conf/mobisys/HuangQGMSS12,DBLP:conf/mobicom/NikraveshGQMS16} demonstrate that round-trip times follow heavy-tailed distributions. We adopt the Pareto distribution as our analytical model:
\begin{equation}
	\mathbb{P}(L > x) = \left(\frac{x_m}{x}\right)^\alpha, \quad x \geq x_m
	\tag{20}
\end{equation}
where $x_m$ is the scale parameter (minimum latency) and $\alpha$ is the shape parameter governing tail heaviness. When $\alpha \leq 2$, variance is infinite; when $\alpha \leq 1$, even the mean diverges. Measurements from the MobiPerf dataset~\cite{mobileperf} and related studies suggest $\alpha \in [1.5, 3]$ for typical mobile networks.

The Pareto distribution satisfies the maximum domain of attraction property for extreme value theory~\cite{dehaan2006extreme}, making it a canonical choice for analyzing tail behavior in distributed systems.

\subsection{Impact on Synchronization Barriers}

At each synchronization barrier in \sys's DAG traversal, the PS waits for all $D$ assigned devices. The barrier completion time is determined by the maximum latency:
\begin{equation}
	T_{\text{barrier}} = \max_{k=1}^{D} L_k
	\tag{21}
\end{equation}

For Pareto-distributed latencies with $\alpha > 1$ (finite mean), the expected maximum scales according to extreme value theory~\cite{leadbetter2012extremes}:
\begin{equation}
	\mathbb{E}\left[\max_{k=1}^D L_k\right] \sim x_m \cdot \frac{\alpha}{\alpha - 1} \cdot D^{1/\alpha}
	\tag{22}
\end{equation}

This scaling is substantially worse than the $O(\log D)$ growth for light-tailed distributions such as exponential or Gaussian. Table~5 quantifies this difference.

\begin{table}[h]
	\centering
	\begin{tabular}{lcc}
		\toprule
		\textbf{Distribution $\alpha$} & \textbf{Expected Max} & \textbf{Expected Max} \\
		                               & D=100                 & D=1000                \\
		\midrule
		Exponential                    & $5.2 \cdot x_m$       & $6.9 \cdot x_m$       \\
		Pareto 3                       & $6.9 \cdot x_m$       & $14.9 \cdot x_m$      \\
		Pareto 2                       & $10.0 \cdot x_m$      & $31.6 \cdot x_m$      \\
		Pareto 1.5                     & $21.5 \cdot x_m$      & $100.0 \cdot x_m$     \\
		\bottomrule
	\end{tabular}
	\caption{Expected maximum latency as a multiple of scale parameter for different tail behaviors.}
\end{table}

\subsection{Tail-Aware Cost Model}

We augment the cost model (Equation~2) to account for tail risk using Conditional Value-at-Risk (CVaR), also known as Expected Shortfall~\cite{rockafellar2000optimization}:
\begin{equation}
	C_{\text{GEMM}}(s,p) = \text{CVaR}_\beta\left[\max\left(C_{\text{COMM}}(s,p), C_{\text{COMP}}(s,p)\right)\right]
	\tag{23}
\end{equation}
where $\text{CVaR}_\beta$ denotes the expected value in the worst $\beta$-fraction of outcomes. For operational relevance in training stability, we recommend $\beta = 0.05$, capturing 95th percentile behavior.

For Pareto-distributed latencies, the CVaR admits closed form:
\begin{equation}
	\text{CVaR}_\beta[L] = \frac{x_m}{\beta^{1/\alpha}} \cdot \frac{\alpha}{\alpha - 1}
	\tag{24}
\end{equation}

Alternatively, the objective can incorporate a variance penalty for risk-averse optimization:
\begin{equation}
	\min_{\{x_{ik}\}} \quad \mathbb{E}[T_{\text{total}}] + \lambda \cdot \sqrt{\text{Var}(T_{\text{total}})}
	\tag{25}
\end{equation}
where $\lambda \geq 0$ controls risk aversion. When latencies are independent across devices, this formulation can be solved via second-order cone programming~\cite{lobo1998applications}, maintaining computational tractability.

\subsection{Straggler Mitigation Guarantees}

\sys's approach of excluding stragglers (Section~3.1) represents one mitigation strategy. We analyze the theoretical tradeoffs of alternative approaches.

\textbf{Speculative Execution.} Assign each row-column pair to $r$ devices and use the first response. The probability that all $r$ copies experience tail latency decreases as:
\[
	\mathbb{P}(L > t)^r = \left(\frac{x_m}{t}\right)^{r\alpha}
\]
The expected completion time for $r$-way replication is:
\begin{equation}
	\mathbb{E}\left[\min_{j=1}^r L_j\right] = x_m \cdot \frac{r\alpha}{r\alpha - 1} \cdot r^{-1/\alpha}
	\tag{26}
\end{equation}

The optimal redundancy factor balances latency reduction against communication overhead. Setting marginal benefit equal to marginal cost yields:
\begin{equation}
	r^* \approx \left(\frac{C_{\text{comm}}}{C_{\text{tail}} \cdot \alpha}\right)^{\alpha/(\alpha+1)}
	\tag{27}
\end{equation}
where $C_{\text{comm}}$ is the per-replica communication cost and $C_{\text{tail}}$ is the cost of experiencing tail latency. For $\alpha = 2$ and moderate tail penalty, this suggests $r^* \in [2, 4]$.

\textbf{Coded Computation.} Rather than full replication, erasure codes enable recovery from any $k$ of $n$ responses~\cite{DBLP:journals/tit/LeeLPPR18}. The makespan becomes the $k$-th order statistic of $n$ latencies. For Pareto distributions:
\begin{equation}
	\mathbb{E}[L_{(k:n)}] \sim x_m \cdot \frac{\Gamma(n+1)\Gamma(1-1/\alpha)}{\Gamma(n-k+1+1/\alpha)\Gamma(k)}
	\tag{28}
\end{equation}
Setting $n - k = O(n^{1 - 1/\alpha})$ yields expected latency of $O(x_m)$ with redundancy overhead $O(n^{-1/\alpha})$, providing a principled tradeoff between communication cost and tail latency mitigation.

\subsection{Implications for \sys Design}

Incorporating fat-tailed latency analysis yields three design insights for \sys deployments.

First, the optimal device count depends on the tail parameter $\alpha$. With heavier tails, marginal benefit from additional devices diminishes faster. The refined scaling condition becomes:
\begin{equation}
	D^* \approx \left(\frac{W_{\text{GEMM}}}{L_{\text{median}} \cdot W_d}\right)^{\alpha/(\alpha+1)}
	\tag{29}
\end{equation}
where $W_{\text{GEMM}}$ is workload per GEMM, $L_{\text{median}}$ is median latency, and $W_d$ is downlink bandwidth. For $\alpha = 2$, this yields $D^* \propto (W_{\text{GEMM}})^{2/3}$ rather than linear scaling.

Second, the PS architecture provides a natural control point for straggler mitigation. The PS can maintain empirical latency distributions for each device using exponential moving averages and dynamically adjust assignments. A Thompson sampling approach~\cite{russo2018tutorial} (suggested future extension, not implemented in current \sys) balances exploration (learning device characteristics) with exploitation (assigning to reliable devices).

Third, bandwidth asymmetry interacts multiplicatively with tail effects. If uplink latencies exhibit heavier tails than downlink—plausible given wireless upload characteristics and contention -- \sys's design choice to minimize uplink communication provides compounding benefits beyond mean-case analysis. Specifically, if downlink has tail parameter $\alpha_d$ and uplink has $\alpha_u < \alpha_d$, the effective benefit of reducing uplink volume by factor $\gamma$ scales as:
\[
	\gamma^{1 + 1/\alpha_u - 1/\alpha_d}
\]

\section{Implementation Details}
\label{sec:appendix-impl-detail}

\mypar{System design choices}
We focus on GEMM operations, particularly in attention and MLP layers, as they are the most computationally intensive, exhibiting cubic time complexity with respect to model parameters, while other operations have quadratic complexity~\cite{megatron,scaling-law}. 
While most work on partitioning GEMM focuses on sparse matrices~\cite{spgemm-part,spgemm-survey}, our approach targets the dense matrices typical in foundation models and addresses challenges unique to edge training.

Our approach uses synchronized communication, with the PS waiting for responses from all assigned workers. The system processes training batches with a fixed optimal batch size globally. It also supports asynchronous training, synchronizing forward and backward passes while allowing asynchronous gradient accumulation.

Managing a large number of devices has been extensively studied in edge orchestration frameworks. Existing systems like KubeEdge~\cite{kubeedge} and Orchestra~\cite{oakestra} support large-scale cloud-edge communication, typically relying on protocols such as MQTT~\cite{standard2019mqtt} or AMQP~\cite{standard2012oasis} for communication.
We adopt these protocols to broadcast rows and columns to specific groups of devices as determined by the solver. Parameter server frameworks, such as FedScale~\cite{fedscale} and LIFL~\cite{lifl}, provide functionality for device management in federated learning contexts. We leverage device registration and keep-alive mechanisms from these frameworks in our system.

\mypar{Solver Implementation}
We use the Gurobi~\cite{gurobi} solver to compute the optimal solution. For the largest tested search space—1024 devices and a model size of 70B—the solver completes in approximately 10 minutes, considering six types of GEMM operations involved in training.
For a given configuration, including the number of devices, device compute capabilities, bandwidth, model batch size, and sequence length, the solver needs to be run only once. Since the training process typically involves thousands to millions of batches, with each batch taking over a minute, the solver's runtime overhead is negligible.

\end{document}